\documentclass[letterpaper,twocolumn,10pt]{article} 
\usepackage{usenix-2020-09}

\usepackage{datetime}
\usepackage{fancyhdr}

\usepackage[utf8]{inputenc}

\usepackage{booktabs}
\usepackage{caption}
\usepackage{subcaption}
\usepackage{dcolumn}
\usepackage{lscape} 
\usepackage{colortbl}

\usepackage{rotating}
\usepackage{multicol}
\usepackage{multirow}
\usepackage{array} 
\usepackage{balance}
\usepackage{url}
\usepackage{svg}
\usepackage{amsmath}
\usepackage{enumitem}
\setlist[itemize]{leftmargin=2em}

\usepackage{tcolorbox}

\usepackage{color} 

\newcommand*\dash{\unskip\kern.16667em---\penalty\exhyphenpenalty
        \hskip.16667em\relax
}

\renewcommand{\paragraph}[1]{\vspace{0.08in}\noindent\textbf{#1.}}

\AtBeginEnvironment{quote}{\itshape}

\fancypagestyle{firstpage}{%
  \lhead{A shortened version of this paper appears in the \textit{Proceedings of the 32th USENIX Security Symposium}, August 2023. This is the full version, available online at \url{https://securityethics.cs.washington.edu} (posted February 2023, revised August 2023); also available at \url{https://arxiv.org/abs/2302.14326}.}
  \rhead{}
}

\title{Ethical Frameworks and Computer Security Trolley Problems: \\ Foundations for Conversations 
}

\author{
Tadayoshi Kohno\\
University of Washington
\and
Yasemin Acar \\
Paderborn University \& \\George Washington University 
\and
Wulf Loh\\
Universit\"at T\"ubingen
}

\begin{document}

\maketitle
      \thispagestyle{firstpage}

\begin{abstract}

The computer security research community regularly tackles ethical questions. The field of ethics / moral philosophy has for centuries considered what it means to be ``morally good'' or at least ``morally allowed / acceptable''. Among philosophy's contributions are (1) frameworks for evaluating the morality of actions \dash including the well-established consequentialist and deontological frameworks \dash and (2) scenarios (like trolley problems) featuring moral dilemmas that can facilitate discussion about and intellectual inquiry into different perspectives on moral reasoning and decision-making. In a classic trolley problem, consequentialist and deontological analyses may render \emph{different} outcomes. In this research, we explicitly make and explore connections between moral questions in computer security research and ethics / moral philosophy through the creation and analysis of trolley problem-like computer security-themed moral dilemmas. In doing so, we seek to contribute to conversations among security researchers about the morality of security research-related decisions. We explicitly do \emph{not} seek to define \emph{what} is morally right or wrong, nor do we argue for one framework over another. Indeed, the consequentialist and deontological frameworks that we center, in addition to coming to different conclusions for our scenarios, have significant limitations. Instead, by offering our scenarios and by comparing two different approaches to ethics, we strive to contribute to \emph{how} the computer security research field considers and converses about ethical questions, especially when there are different perspectives on what is morally right or acceptable. 
Our vision is for this work to be broadly useful to the computer security community, including to researchers as they embark on (or choose not to embark on), conduct, and write about their research, to program committees as they evaluate submissions, and to educators as they teach about computer security and ethics.

\end{abstract}

\section{Introduction}

We believe in the essentiality of maintaining high ethical standards when conducting and evaluating computer security research.
As examples of the field's\footnote{When we say ``the field'', we refer to the computer security research field even though our team is composed of both computer security researchers and a moral philosopher. We use this terminology because our primary goal is to contribute to the computer security research field.} commitment to moral considerations, conference program committees are leveraging ethics review boards and authors are discussing ethics in submissions. There also exist  tools to help community members make adequate moral decisions, such as the 2012 Menlo Report~\cite{Menlo} and  recent author guidelines in security conference calls for 
papers, e.g.,~\cite{Oakland2023CFP,Oakland2022CFP,USENIXSecurity2023CFP}.

However, challenges still arise. Central to these challenges is that in some cases there may \textit{not} be universal agreement on what constitutes an adequate decision.
Consider, for example, a hypothetical scenario (our Scenario~A) in which researchers find a vulnerability in a wireless implantable medical device. Assume that the device manufacturer is out of business and, hence, it is impossible to patch the vulnerability. Also, assume that there is \emph{zero} chance of the vulnerability ever being exploited,
even if adversaries know about it.\footnote{To enable us to focus on the philosophical aspects of ethics and morality and not become entangled in real-world details, we make simplifying assumptions in this and all our scenarios. We elaborate on this decision in Section~\ref{sec:scenarios-initial}.} 
Should the researchers \textit{disclose the vulnerability} to the government and the public, thereby respecting patients' right to be informed (a key component of the ``respect for persons'' principle of the Menlo Report~\cite{Menlo} as well as the earlier Belmont Report~\cite{Belmont} and the principle of ``autonomy'' in the Principles of Biomedical Ethics~\cite{BeauchampChildress})? Or, should the researchers, knowing that adversaries would \emph{never} manifest but that a knowledge of the vulnerability's existence could harm patients (who might remove the device from their bodies and hence lose the health benefits out of unnecessary concerns), \textit{not disclose} the vulnerability to the government and the public (thereby respecting the principle of ``beneficence'' and the avoidance of harm, another core element of
the Menlo Report~\cite{Menlo}, the Belmont Report~\cite{Belmont}, and the Principles of Biomedical Ethics~\cite{BeauchampChildress})?

There are strong arguments for \emph{both} decisions.
In a situation with conflicting arguments, how are we as a field to make the right decision?
We argue that whatever process is used,
that process will benefit from being informed by philosophy's understanding of the different approaches that people take to ethics \dash approaches that \textit{can} result in different people coming to different conclusions.
Hence, our research.
In short, we seek to contribute to future conversations about \emph{what} is morally right, good, or allowed, and we do so by studying \emph{how}, from a philosophical perspective, to have such discussions.

\paragraph{Ethics and Moral Philosophy}
The field of ethics / moral philosophy centers the question of what it means to be ``good'', ``morally right'', or ``morally allowed'' (i.e., not prescribed but also not forbidden). Even in the field of philosophy, there is no consensus for \textit{what} this exactly means. But, there is a mature understanding for \emph{how} to discuss what it might mean. 
Conversations about ``good'' and ``bad'' begin by centering a perspective \dash an \emph{ethical framework}.

\paragraph{Ethical Frameworks} Ethical frameworks define approaches for reasoning about what is morally right or wrong, i.e., what is ``good'' or ``bad''.
Two of today's leading frameworks (or, more precisely, categories of frameworks) are \textit{consequentialist}  and \textit{deontological ethics}. Consequentialist ethics centers questions about the impacts (consequences) of different decisions. Under consequentialist ethics, one might assess the benefits and harms of different options before making a decision that maximizes net benefits. Deontological ethics centers questions about duties (deon) and rights. Under deontological ethics, one might ask what rights different stakeholders have, e.g., a right to privacy or a right to autonomy.

We center consequentialist and deontological ethics \dash and in particular utilitarianism  and Kantian deontological ethics, respectively (Section~\ref{sec:frameworks}) \dash
in our study because of (1) their prominence in the field of ethics / moral philosophy (they are two of the three leading frameworks) and (2) their existing impact on the computer security research field's approach to ethics and morality (e.g., the Menlo Report~\cite{Menlo} derives from the Belmont Report~\cite{Belmont}, which itself embeds both consequentialist and deontological elements). We stress, however, that both consequentialist and deontological ethics have limitations and that by centering them we are \emph{not} arguing that anyone adopt a strict consequentialist or  deontological perspective. At a minimum, one might include considerations from both frameworks, as the Menlo Report~\cite{Menlo} does. As we discuss more in Section~\ref{sec:scenarios-initial}, modern frameworks include a more critical perspective. Additionally, much of philosophy's discussion of consequentialist and deontological ethics centers a Western perspective. While Western frameworks encompass ethical considerations that are part of non-Western traditions (e.g., about duties towards each other, the nature of fundamentally relating to each other, the outcomes of actions / policies, and so on), each tradition has its own unique history and elements. Although outside the scope of this work, we encourage the computer security research community to gain greater familiarity with other frameworks as well.

\paragraph{Our Work} 
As exemplified by the Menlo Report~\cite{Menlo} and recent calls for papers, 
e.g.,~\cite{Oakland2023CFP,Oakland2022CFP,USENIXSecurity2023CFP}, 
there are \emph{already} connections between the computer security research field and ethics / moral philosophy. Our assessment is that many of these connections are implicit. 
We seek to make these connections explicit and, by doing so, contribute to \emph{how} the field discusses and considers moral questions.

To do our research, we composed a team of researchers consisting of both those  trained in  computer security research  \emph{and} those  trained in moral philosophy. The computer security researchers on our team have significant prior work addressing and discussing ethical questions in computer security research. However, prior to this collaboration, the security researchers approached ethics from a ``we should be good'' and ``having thought carefully and talked with others, I think this is right'' approach rather than from an approach informed by ethics / moral philosophy. The moral philosopher on our team has significant prior work in applied ethics outside of computer security. Our work is thus cross-disciplinary and could be read as both a work in philosophy (particularly normative and applied ethics) and (we believe) a contribution to the computer security research field.

\paragraph{Goals, Methods, and  Findings}
We seek to leverage tools and insights from ethics / moral philosophy to facilitate clear, thoughtful, and rigorous conversations within the computer security research field about what are morally right or allowed decisions / policies / institutions \dash i.e., we seek to contribute to \emph{how} the field discusses moral questions. 
We do \textit{not} seek to  define \textit{what} (morally) ``right'' or ``good'' means (which would be a metaethical question).

Our methodology is to:
\begin{itemize}
    \item Develop computer security scenarios reminiscent of classical ethical dilemmas and for which evaluations under different ethical frameworks justify different outcomes; our scenarios are akin to philosophy's classic trolley problems, which we describe later. (Section~\ref{sec:scenarios-initial}.)
    \item Explore those computer security scenarios using both consequentialist (utilitarianism) and (Kantian) deontological analyses. (Section~\ref{sec:analysis-scenarios-initial}.)
    \item Develop additional computer security scenarios that, individually, may not pose ethical dilemmas (i.e., there may be stronger agreement for what constitutes a morally right or allowed decision for some scenarios) but that, together, facilitate deeper explorations about moral considerations within the field. 
        (Summarized in Section~\ref{sec:scenarios-more}.)
    \item Reflect upon the above scenarios and explorations and derive lessons about \emph{how} to have informed conversations about ethics and morality in the computer security research community. (Section~\ref{sec:discussion}.)
\end{itemize}

\paragraph{Summary of Our Three Main Scenarios} We summarized Scenario~A above. Scenario~B explores the morality of studying stolen data \dash data that people did not intend to be public. Scenario~C explores what to do if a program committee member encounters a submission containing undisclosed information about their company's product.
All our scenarios are based on actual situations encountered within the computer security research community, though we modified the scenarios to make them more conducive to ethical analyses, per our research goals. 
While reflective of real-world scenarios, our scenarios do not cover the full spectrum of moral dilemmas encountered within the security research community, nor is it our intent to do so.

\paragraph{Example Use Case of Our Results: Program Committee Discussions}
The security researchers on this team have on multiple occasions encountered the following situation: 
\begin{itemize}
    \item A paper is submitted to a peer-reviewed conference. The paper reports on work that one program committee member flags as possibly unethical.
    \item Program committee members discuss the morality of the work but cannot agree; 
    some committee members think it was ethical, and others think it was not.
\end{itemize}

Such disagreements can be challenging if, for example, some committee members adopt consequentialist perspectives and other committee members adopt deontological perspectives, but the committee members do not realize that they are using different frameworks for evaluating morality.
Prior to this collaboration, we (the security researchers on this team) did not have the tools and language to untangle such  disagreements.
Now, through this collaboration and the exploration of computer security scenarios via the consequentialist and deontological frameworks, we have that language.

\paragraph{Example Use Case of Our Results: Discussing Research Path} 
In many cases, there may already exist clarity for researchers on how to navigate moral questions, e.g., researchers might follow the recommendations in the Menlo Report~\cite{Menlo}.
However, there may remain times when clarity does not exist, e.g., when there are tensions between what is morally right from a benefits / harms perspective (consequentialist ethics) and what is right from a duties / rights perspective (deontological ethics).
Through the articulation of established ethical frameworks, and through the exploration of computer security scenarios via these frameworks, we hope to help researchers have more methodical and informed discussions about ethics and morality when there are such tensions.
Since different frameworks can lead to different conclusions of what is morally right, however, we stress that the frameworks should \emph{not} be used to justify a path that researchers have \emph{a priori} decided that they want to take. Rather, we argue that the ethically correct process is to \textit{center} ethics in the decision of whether or not to do a research project or do some component of the research and accept that sometimes the answer is ``no''.

\paragraph{Example Use Case of Our Results: Community Conversations and Education}
In ethics / moral philosophy, ethical dilemmas, like the trolley problems in Section~\ref{sec:background:trolley}, often feature prominently in education and also certain debates within the community. Consider, for example, the centrality of trolley problems in the popular book \emph{Justice} by Sandel~\cite{Justice}. We believe that our computer security dilemmas (Section~\ref{sec:scenarios-initial}) and other scenarios 
(Section~\ref{sec:scenarios-more}) 
can likewise facilitate community conversations and education. For example, conference ethics review committees (and the field at large) could use these scenarios as starting points for discussing norms, and instructors could use these scenarios to help students understand ethical thinking in the computer security field (and the importance of ethics in the first place). Additionally, we hope that our work inspires others to likewise create and share computer security-themed ethical scenarios and trolley problems for broader community conversations and educational discussions.
As we create  additional scenarios and educational materials, we will put them online at \url{https://securityethics.cs.washington.edu/}.

\paragraph{Publication Information}
A conference version of this paper appears in the Proceedings of the 32th USENIX Security Symposium, August 2023~\cite{SecurityEthicsConference}.
\section{Motivation and Background}
\label{sec:background}

We begin in Section~\ref{sec:background:ethics} with a brief background on ethics / moral philosophy and then turn to a background discussion on ethics and computer security research in Section~\ref{sec:background:security}. In Section~\ref{sec:background:trolley} we summarize the trolley problem, a classic moral dilemma. 

\begin{figure*}
\begin{tcolorbox}
\begin{center}

\textbf{A Classic Dilemma: The Trolley Problem}\end{center}
\textbf{Context:}
\begin{itemize}[itemsep=0pt, parsep=2pt]
\item A runaway trolley with no brakes is heading straight along a set of tracks.
\item Five people are tied to those tracks.
\item One person is tied to an alternate set of tracks.
\item A trolley operator has the ability to change the trolley's path and make it head down the alternate set of tracks.
\end{itemize}

\textbf{The choice for the trolley operator:}
\begin{itemize}[itemsep=0pt, parsep=2pt]
    \item \textit{Do nothing:} Five people die.
    \item \textit{Make the trolley take the alternate set of tracks:} One person dies.
\end{itemize}
\end{tcolorbox}
\caption{The trolley problem is a classic thought experiment / ethical dilemma.}
\label{fig:scenario:trolley}

\end{figure*}

\subsection{Ethics / Moral Philosophy}
\label{sec:background:ethics}

Ethics / moral philosophy is a field that has existed for centuries. In Western culture, the most well-known ethical frameworks are virtue ethics (most notably developed by ancient Greek philosophers such as Plato and Aristotle), deontological ethics (a famous example from German Enlightenment philosopher Immanuel Kant), and utilitarianism (an example of consequentialist ethics, first developed by Jeremy Bentham and John Stuart Mill). In other cultures, classic ethical frameworks include Confucianism, Daoism (as first coined by Laozi), and Ubuntu.

As a field that is centuries old and spans cultures and histories, it is natural that there is no universal consensus on what, precisely, ethics and morality mean. For our work, we use \textit{ethics / moral philosophy} to refer to the (scientific) exploration of \emph{how} to consider, evaluate, and discuss moral questions,\footnote{Whereas ``moral philosophy'' refers to a mainly philosophical endeavor,  ``ethics'' also comprises non-strictly-philosophical (e.g., theological) reasoning.} and \textit{morality} to refer to the object of this exploration. For example, ethicists / moral philosophers use ethics in the sense of moral reasoning to determine whether an action, social institution, or  set of norms is moral or not~\cite{sep-morality-definition}.

Modern ethicists often use \textit{ethics} and \textit{morality} interchangeably.\footnote{Against this, Habermas argues that ``ethics'' refers to questions about the good life, whereas ``morality'' is concerned with what we owe others~\cite{habermas1994justification}.}
Thus, the computer security field is not wrong in its use of the term \textit{ethics} to encompass both ethics and morality. When precision is not necessary, we may do so as well.

We provide a deeper background on ethical frameworks in Section~\ref{sec:frameworks}.

\subsection{Ethics and Computer Security Research}
\label{sec:background:security}

While Institutional Review Boards at U.S.\ universities reference the 1976 Belmont Report~\cite{Belmont}, they often only have the capacity to evaluate the direct treatment of research subjects, and often work across biomedical sciences, with computer security research not their primary focus. European universities and other research institutions may reference general ``good scientific practices'', and may discuss dual use of research results (civilian and military), but often omit formal ethics reviews of research ideas, leaving the ethical decisions to researchers. 

The security community has a longstanding tradition of discussing topics of ethical concern, even if a direct connection to ethics and morality is not always present. Early in the evolution of the modern computer security field, there were the ``crypto wars'', which centered debates over cryptographic strength, export controls, people's need for strong cryptography versus the illicit use of cryptography, and law enforcement access to encrypted data, e.g.,~\cite{blaze1994protocol,PhilZimmermannLetter,CryptoWars}. Echos of these debates continue today~\cite{AcademiesCrypto}. 
Another early debate was on whether strong cryptographic mechanisms should be patented and only commercially available, or whether strong cryptographic mechanisms should be freely available~\cite{SimsonWebSecurity}. 
Phillip Rogaway explicitly discusses morality and cryptography in his paper entitled ``The Moral Character of Cryptographic Work''~\cite{rogaway2015moral}.

The computer security community also long wrestled with when and how to disclose computer security vulnerabilities to product maintainers and the public. Although there are now general best practices for coordinated vulnerability disclosure, those best practices do not always capture the full complexity that researchers and industry must consider, as recently exemplified by the complexities faced during the Spectre and Meltdown disclosure process~\cite{AcademiesSpectre} or the lawsuit preventing the publication of a paper at USENIX Security 2013 (eventually appearing as a supplement in 2015~\cite{Megamos,VWarticle}).
Additionally, the history of coordinated vulnerability disclosure \dash and its evolution from so-called responsible disclosure \dash is itself a microcosm of ethical considerations, as the term ``responsible disclosure'' centers one perspective on what constitutes responsible behavior~\cite{KatieTweet,KatieBlog}. When disclosing vulnerabilities across community boundaries, one may also encounter ideological differences regarding the morality of public vulnerability disclosures, as exemplified by Matt Blaze's experience after publicly disclosing vulnerabilities in physical locks~\cite{BlazeTweet,BlazeLocks}.

There are also longstanding ethical considerations with respect to networks and systems-related computer security research.
For example, there is a tradition of considering the morality of anonymous and censorship-resistant communications, e.g., with Free Haven~\cite{FreeHaven} and then later Tor~\cite{Tor}.
Because research instruments that scan the Internet have the potential to unintentionally cause network-attached devices to crash or create additional work for network maintainers (who might interpret the network scans as active attacks), the research community has developed ethical practices for Internet scanning, e.g.,~\cite{durumeric2013zmap}. 
The field has also explored the ethics (and legality) of using network trace data in research~\cite{NetworkLegal}. 
The field has further wrestled with (and answered) questions such as: is it ethical to include researcher computers in botnets, in order to study those botnets, e.g.,~\cite{holz2008measurements,Spamalytics,UCSB:Torpig}? 
And, using such a  botnet infrastructure, how to ethically ``send'' spam in order to study the spam ecosystem~\cite{Spamalytics}. Researchers have additionally asked whether it is okay for researchers to clean up computers infected by botnets~\cite{BotnetEthicsProblems}; although~\cite{BotnetEthicsProblems} does not make explicit connections to ethical frameworks, it, like our work, explores questions for which there are  valid affirmative and negative arguments.

Turning to the most recent decade, 
a set of surveys in 2014 assessed crowdworkers' reactions to (at the time) recent controversial studies, including some studies mentioned here~\cite{StuartSurvey1,StuartSurvey2}.
A 2017 paper analyzed the ethics of using stolen  password (or other) datasets in academic research~\cite{thomas2017ethical}; researchers feared ``worse'' data from using research-study-generated passwords compared to real-world password sets, but were wary of the ethical implications of publishing research that benefited from illicit datasets.
After the publication of Internet censorship research that might have caused browsers of unsuspecting users to contact possibly censored domains, researchers in 2015 discussed the ethics of measuring censorship~\cite{ethicscensorship}.  In another discussion-provoking example, researchers pushed vulnerable commits to open source projects to ``test'' whether important projects were vulnerable to such attacks, an act that was deemed unethical by both the open source community and the research community; the paper was ultimately withdrawn by the authors~\cite{holz2021ieee}. Interestingly, here, security researchers and their IRB had missed the connection to human subjects research, while also misjudging how their experiment on live projects would be perceived. These are, of course, just some examples of ethical discussions continuing to this date. 

In part in response to the types of issues summarized above, in 2012 the security community developed the Menlo Report~\cite{Menlo}, which centers the Belmont Report's principles of justice, beneficence, and respect for humans (not just research subjects), and applies them to computer security research, highlighting that for this specific field, instead of ``only'' caring about subject rights, research may be done on data and live systems and may impact computer systems and their users beyond consenting participants. The report suggests transparency, ethics review, and careful ethical considerations.
The Menlo Report~\cite{Menlo}, as an applied ethics framework, lowered the barrier of entry to ethical practices \dash those applying the principles in the Menlo Report do not need to become ethicists.
The past several years have seen an uptick in recognition of the value and significance of the Menlo Report~\cite{Menlo}, e.g., the explicit reference to the Menlo Report in the 2022 and 2023 IEEE Symposium on Security \& Privacy and the 2023 USENIX Security call for papers~\cite{Oakland2023CFP,Oakland2022CFP,USENIXSecurity2023CFP}.
Our premise is that the application of the Menlo Report~\cite{Menlo}, and security ethics conversations in general, can be further enriched with a greater understanding of ethics / moral philosophy, thus this work.

The intersection between ethics and research is expansive, and there are many important contributions across computer science, and science and technology in general. We survey a few additional contributions here, though acknowledge that our survey is not nearly sufficient. Chivukula et al.\ provide a survey of ethics-focused design methods~\cite{chivukula2021surveying}. An example method is value sensitive design~\cite{friedman2007human,friedman2006development,friedman2019value}. The entire field of  Science and Technology Studies places technologies in the context of people and society and hence deeply integrates considerations of ethics. 
Examples of ethical considerations in other subfields of computer science include biases in natural language processing, e.g.,~\cite{caliskan2017semantics, field2021survey,d2022ethics}, in computer vision technologies, e.g.,~\cite{buolamwini2018gender,raji2020saving}, and the use of ``public'' data in participant studies, e.g.,~\cite{bruckman2002studying,FieslerTwitter} (see also Section~\ref{sec:scenario:immoraldata}). Indeed, entire conferences exist with foci related to ethics, e.g., the ACM Conference on Fairness, Accountability, and Transparency and the AAAI/ACM Conference on Artificial Intelligence, Ethics, and Society.

Computer security venues introduced ethics statements into their calls for papers only within the last decade; some have since added ethics review task forces to the peer review process~\cite{ethicsinsecurityresearch}. 
The Menlo Report~\cite{Menlo}, as well as past experiences within the community, have strongly influenced the community's current approach to ethics. 
There continue to be numerous points of discussion. For example, who should make ethics-related decisions on whether to do or publish research? What is the role of the security research community at large, including but not limited to decisions during peer review? And what are we, as a community, trying to enable or prevent? Discussions have included reasonings such as ``authors made ethics mistakes so should not be published''; ``the scientific contribution merits publication with a disclaimer that this methodology should not be repeated''; and ``we can prevent future harms by publishing, with ethics discussions, papers detailing experiments that should never have been done''. Among all these, the discussion includes observations that allowing papers to be published through peer review may create incentives for immoral research.

\subsection{A Classic Moral Dilemma}

\label{sec:background:trolley}

Ethicists / moral philosophers have, for generations, proposed dilemmas for ethical debate and consideration. A classic dilemma (or, more precisely, family of dilemmas) are the ``trolley problems''. These are dilemmas because they present a choice between two options, both  of which contain undesired aspects. Therefore, different ethical frameworks potentially present different answers to such dilemmas. Some authors (among them Philippa Foot herself, who came up with the original trolley problem~\cite{Foot2002MoralDilemmas})
take them to show that people's moral intuitions will most likely diverge in important cases.

Figure~\ref{fig:scenario:trolley} presents an archetypical trolley problem. In this trolley problem, a runaway trolley with no brakes is heading straight down a track. Five people are tied to that track. A trolley operator is watching the trolley. They could do nothing, in which case five people would die. The trolley operator could, however, choose to redirect the trolley down a second, adjacent track. If the operator does so, then the trolley would kill only one person\dash the person tied to that adjacent track.

Philosophers and psychologists have studied people's responses to trolley problems such as in Figure~\ref{fig:scenario:trolley} and, indeed, there is no universal consensus for what constitutes the morally correct action of the trolley operator~\cite{Foot2002MoralDilemmas}. 
In psychology studies, for example, differences can arise due to the moral intuitions and values of the participant and may vary by culture, e.g.,~\cite{conway2013deontological, conway2018sacrificial, gold_colman_pulford_2014, ChineseandWesternersRespondDifferentlytotheTrolleyDilemmas, yamamoto2020causes, lamont2017bridging}. 

Variants of the trolley problem feature different outcomes and can elicit different thought processes and decisions.\footnote{An interactive exploration of different trolley problems is available at \url{https://neal.fun/absurd-trolley-problems/}.} As an example variant, the single person on the alternate track might be a young child whereas the five people on the main track might already be near death. 
In this variant, if the operator does nothing, five near-death people will die; if the operator changes the trolley's track, a single young child will die. 
As another variant, the five people tied to the main track might have tied themselves there intentionally whereas the single person on the other track might be there against their will. Or, the five people on the main track might have been convicted of war crimes by an international tribunal whereas the person on the alternate track is known to have led a virtuous life.

\section{Computer Security Trolley Problems}
\label{sec:scenarios-initial}

We first describe our scenario generation process (Section~\ref{sec:scenarios-initial:process}) and then present our three scenarios (Sections~\ref{sec:scenario:medical}, \ref{sec:scenario:immoraldata}, and~\ref{sec:scenario:inadvertentdisclosure}).

\subsection{Scenario Generation Process}
\label{sec:scenarios-initial:process}

Our research team used a collaborative and interactive process for scenario generation. 
After discussing our initial approach, we present our final methodology and scenario selection criteria.

\paragraph{Initial Approach}
Initially, the security researchers on the team created scenarios representative of scenarios that we (the security researchers) had previously encountered (e.g., as program committee members or as researchers). Our team generated dozens of such scenarios with an initial goal of exhaustively and systematically surfacing the full spectrum of ethical considerations encountered within our field.
While the generative process was important toward normalizing an understanding of scenarios and moral issues in the field across the entire team (including both the security researchers and the philosopher), these scenarios had several key limitations:

\begin{itemize}
    \item \textbf{Too late.} 
    Some scenarios were framed as ``a program committee member reads a conference submission in which the authors did such-and-such; the program committee member believes that such-and-such should not have been done; should the program committee accept the paper?'' 
    \item \textbf{Open-ended.} Other scenarios were very open-ended and highly unconstrained, e.g., ``here is an issue that a research group encountered, what should they do?''
    \item \textbf{Not a dilemma.} Some scenarios had relatively clear and uncontroversial moral implications;  we encountered them as program committee members because (for example) of an oversight by the authors of a paper submission, e.g., because the authors assumed that the IRB process was sufficient to cover all aspects of moral decision-making.
    \item \textbf{Indecisive.} Some of our scenarios did not have conclusive decisions under different ethical frameworks, at least not without significant additional information that would greatly expand the scenarios and make them unwieldy.
\end{itemize}

The ``too late'' scenarios all shared a common theme: researchers made decision $X$, for some $X$; what should the program committee do if they question the morality of $X$?
A discussion of the ethical processes for program committees when encountering such papers is important, and indeed we consider a family of such scenarios in Section~\ref{sec:scenarios-more}. 
For our core ethical dilemmas (this section), we sought scenarios featuring a decision
\emph{before} a controversial act $X$ is committed in the first place.
As a concrete example, for our Scenario~B (to be described), an initial version featured a scenario in which a program committee reviews a paper that studies data that some program committee members believe should not have been studied. What should the program committee do? Our final Scenario~B asks: should researchers study that data?

The ``open-ended'' scenarios, while representative of what researchers might encounter in the real world, made analyses of the scenarios under established ethical frameworks too unconstrained for focused treatments. 
As evidenced by our team's internal discussions, when faced with open-ended questions of the form ``what should the researchers do?'', it is possible to spend hours, and hence volumes of written pages, exploring different possible paths forward. While for some scenarios such explorations would be important contributions of their own, those are not the contributions we sought with this work. Rather, we wanted scenarios conducive to short, precise, and focused analyses with minimal (binary) options.

The ``not a dilemma'' scenarios were intellectually interesting and important in establishing our team's shared understanding of questions of morality and computer security research. However, because these scenarios were not actual dilemmas, evaluation under different ethical frameworks resulted in the same conclusions and hence were not as generative of philosophical explorations as scenarios that yielded different conclusions under different ethical frameworks. In short, we sought scenarios for which people \dash including computer security research community members \dash might, through sound  reasoning, plausibly disagree.
Still, our team found value in comparing multiple ``not a dilemma'' scenarios: even if two scenarios do not individually present dilemmas, the comparison of what is morally correct in those two scenarios can contribute insights into how our field reasons about moral questions. 
We return to the comparison of multiple  scenarios in
Section~\ref{sec:scenarios-more}.

The ``indecisive'' scenarios featured decisions for which all possible choices would result in ``comparable'' benefits / harms that would need extensive empirical work to assess. In the real world, if one were to encounter such a situation, a significant portion of the conversation might center on assessing those empirical claims. For our work, we wanted to center ethical and moral thought processes, not empirical questions. Hence, we sought scenarios without complicated benefits / harms calculus.

\paragraph{Revised Approach: Criteria, Creation, and Validation} \label{sec:scenarios-initial:criteria}
Informed by the results of our analyses of and conversations about our initial scenarios, our team developed the following criteria for scenario generation:
\begin{itemize}
    \item \textbf{Early.} We sought scenarios that featured moral questions that actors (e.g., researchers) might encounter about their own future actions, not questions about what to do after it has been determined that researchers have already committed a morally questionable act. 
    \item 
    \textbf{Binary options.} We sought scenarios that \dash like the trolley problems \dash have binary options for some actor (e.g., the trolley operator in the trolley problem in Figure~\ref{fig:scenario:trolley} or a research team in computer security-related scenarios).
    \item \textbf{Dilemmas.}
    We sought scenarios that were true dilemmas. Specifically,  we sought  scenarios for which analyses under consequentialist and deontological ethics would yield different conclusions.
    \item \textbf{Decisive.} We sought scenarios for which analyses under the consequentialist and deontological   ethical frameworks were clear, straightforward, and decisive. Sometimes this came at the cost of  simplifying and artificially contrasting the ethical traditions to bring out key differences in perspective and focus.
\end{itemize}

Having these criteria meant that our resulting scenarios were not ``too late'', were not  ``open-ended'', were actually dilemmas, and were analyzable in a contrasting way under at least the 
consequentialist and deontological  ethical frameworks. We discuss the consequentialist and deontological  ethical frameworks in Section~\ref{sec:frameworks}. Although we use the term ``early'', we observe that every decision is influenced by earlier, preceding decisions, and hence there may exist important-to-consider scenarios even earlier in a timeline.

Our research team iterated extensively on the creation of scenarios that satisfied these criteria, over regular meetings throughout late summer and fall 2022 and early 2023. Our iteration was both at a high level, focusing on the scenario's overall setup and context, and at a low level, focusing on fine nuances and details. As we iterated on these scenarios, we presented variants in university seminars (at other universities) and in courses (at the undergraduate and graduate levels). After each presentation, we reflected  upon and revised the scenarios as needed to address ambiguities or clarify key aspects relevant to the scenarios' intended moral questions. 
We additionally shared our scenarios with others in the computer security research community for feedback.

In addition to being instrumental to the process of scenario creation, this iterative process also served as scenario validation. Specifically, the iterative process with 
systematic philosophical analyses and external discussions
helped us validate that our scenarios met our ``dilemmas'' and ``decisive'' criteria.  (That our scenarios met the ``early'' and ``binary options'' criteria was easy to assess by construction.)

\paragraph{On the Chosen Scenarios}
Our final scenarios were inspired by scenarios previously encountered by the security field, though we modified the scenarios to meet our design criteria.
While we initially intended for our  scenarios to capture, systematically, the full spectrum of ethical questions that we have encountered within the field, we soon realized that a full analysis of archetypical examples of \emph{all} such questions would be beyond the scope of a single paper, and especially so for a paper (such as this one) intended to dive into and explore the relationship between ethical frameworks and computer security scenarios. Thus, we chose to focus on three scenarios \dash scenarios that are each different from each other and that enable  philosophical analyses of the form we sought.

Our selected scenarios reflect ethical scenarios encountered within our field: what to do after discovering a vulnerability (Scenario~A), whether to study stolen data (Scenario~B), and what to do if a program committee member learns about an undisclosed vulnerability in their company's product (Scenario~C).
We provide additional scenarios in
Section~\ref{sec:scenarios-more}, 
though even then, our full set of scenarios are not exhaustive.
Researchers seeking to create additional scenarios might draw inspiration from their own experiences or from the works surveyed in 
Section~\ref{sec:background:security}.

\paragraph{Based on Reality, But Not Real}
We stress that although our final scenarios are based on reality, they are \emph{not} realistic. Real-world scenarios generally do not present only a binary option to decision-makers \dash they present a medley of options. 
Additionally, to enable precise analyses under different ethical frameworks, our scenarios minimize uncertainty. The real world, on the other hand, is full of uncertainty, e.g., uncertainty about when or if an adversary might manifest or the actual benefits / harms of a technology or exploit. Thus,  assessing benefits / harms (for consequentialist ethics) and rights violations (for deontological ethics) is significantly more challenging in the real world than in our scenarios. Real-world scenarios may have multiple actors simultaneously making decisions, each of which might impact the other actors; in our scenarios, we consider only a single decision-maker. Additionally, to simplify our analyses, we reduce the impacts of decisions on the decision maker in our core scenarios (Scenarios~A, B, and~C); 
we add such impacts back in Scenarios~D$^*$ and~F (Section~\ref{sec:scenarios-more}). 
In the real world, decision-makers may involve others in the decision-making process; our scenario descriptions do not preclude such discussions but leave the final decision in the hands of the specified decision-maker rather than allow for the transference of the decision responsibility to another entity (e.g., a committee or  government).

\paragraph{The Structure of a Scenario}
For each scenario, we use a structure similar to Figure~\ref{fig:scenario:trolley} for the trolley problem. Each scenario centers a decision-maker and has:
\begin{itemize}
    \item \textbf{Context:} The ``context'' of the scenario provides the background context for the decision that the actor needs to make.
    \item \textbf{Choice:} The ``choice''  of the scenario describes two options that the actor must choose between.
\end{itemize}

In the body of the paper, we use prose to describe the context and choice. Appendix~\ref{ap:figures} provides figures, like Figure~\ref{fig:scenario:trolley}, for each of our scenarios. We defer the figures to the appendix because it is not necessary to read the figures in order to understand this paper. Still, the figures provide self-contained descriptions of each scenario and as such may be useful in other contexts (e.g., presenting a scenario  to a class or for discussion with other researchers).

\paragraph{Reflection}
The significant number of scenarios that we initially created that did not meet all our criteria nonetheless provided valuable insights for the formulation of the above criteria. Here, the interdisciplinary work proved especially fruitful in realizing and establishing these criteria. 
We encourage future researchers at the intersection of ethical frameworks and computer security to also adopt or extend the above criteria for scenario creation.

\subsection{Scenario A: Medical Device Vulnerability}
\label{sec:scenario:medical}

Scenario~A centers around researchers who discover a vulnerability in a wireless implantable medical device. 
For a self-contained description of Scenario~A, see Figure~\ref{fig:scenario:medical_device} in Appendix~\ref{ap:figures}.

\paragraph{Context} 
Researchers found a vulnerability in a wireless implantable medical device made by a manufacturer that is no longer in business. Existing patients still use the device and new patients are still receiving the device. It is not possible to update the software on the device and patch the vulnerability. Even if the researchers disclose the vulnerability to the public, there is zero probability of the vulnerability being exploited in the wild. There are no field- or industry-wide gains to be made via the public disclosure and discussion of the vulnerability, e.g., the public disclosure of the vulnerability would not teach the field any new lessons about computer security and medical devices.

\paragraph{The Choice} For this scenario, a disclosure to some sufficiently large group (e.g., all healthcare professionals who work with the relevant medical condition) would eventually result in a disclosure to the public (through information leakage). Hence, the researchers must choose between not disclosing the vulnerability to anyone or disclosing the vulnerability to the government, the healthcare industry, 
and the public.

If the researchers disclose the vulnerability to the public, then patients may be harmed  psychologically (a fear of having a vulnerable / imperfect device even if the likelihood of it being compromised is zero) or physically (the device increases a person's life by ten years; if a patient removes or does not receive the device, they would not receive the health benefits).

If the researchers do not disclose the vulnerability to anyone, then patients do not have the option to make an informed choice with respect to whether they keep the device or, for new patients, whether or not they receive the device.

\paragraph{On this Scenario}
In 2008, one of us (T.K.) co-authored a study that discovered and reported on vulnerabilities in a wireless implantable medical device~\cite{HalHeyRan2008ICD}.
We thought deeply about ethics and responsible disclosure at the time of that study, and the medical device security field has continued to reflect upon ethics and responsible disclosure thereafter, e.g.,~\cite{FDAadvisoryIMD,IMDguide}. 
We designed Scenario~A to center patient-focused elements of consideration: the fundamental rights that patients have and the benefits and harms to patients with either disclosing or not disclosing a vulnerability. 
To center the ethical considerations on the patients, in Scenario~A it is not possible to update the software on the medical device, and hence a traditional coordinated disclosure process of first notifying the manufacturer and then giving them time to respond is not an option (a situation which, unfortunately, is plausible~\cite{IEEESpectrumEye}). Additionally, the healthcare industry has already internalized the importance of computer security for wireless implantable medical devices, e.g., \cite{FDAMITREPlaybook2018,FDAsecurityWeb}, and hence there are no significant field-wide positive impacts with a public disclosure. To meet our scenario design criteria, this scenario  presents only two  options to the researchers. In a real-world scenario, we anticipate much greater involvement from organizations like the U.S.\ Food and Drug Association (FDA); the researchers might even cede the final decision to the FDA or another entity such as U.S.\ CERT.
Additionally, factors such as FDA policy might impact the plausibility of Scenario~A.

\subsection{Scenario B: Studying Stolen Data}
\label{sec:scenario:immoraldata}

Scenario~B centers researchers who are trying to decide whether or not to study stolen data. 
For a self-contained description of this scenario, see Figure~\ref{fig:scenario:immoral_data_jobs} in Appendix~\ref{ap:figures}.

\paragraph{Context} Company B offers a service that matches job applicants with jobs. The public believes that Company~B's AI matching system has racial and gender biases. Some people also believe that Company~B's AI system could be manipulated by adversaries. Adversaries compromise Company~B's servers and steal the entirety of their data, including all data about all past job postings, all past job application packets, and the outputs of all past job-applicant matches from Company~B's AI system. The adversaries also steal all internal details of Company~B's AI matching system, including the underlying ML model. 
The thieves post the stolen material online; a research group obtains a copy of the stolen material as soon as it is publicly available.
Subsequently,
many victims of the data breach \dash the job applicants \dash publicly state their desire for the stolen data to be permanently deleted, everywhere;
all publicly-available copies are then deleted.

\paragraph{The Choice} The research group wishes to study the stolen data and scientifically assess whether Company~B's AI matching system is, in fact, biased. If it is biased, the researchers seek to measure past impacts of those biases, e.g., by counting the number of applicants not forwarded to employers because of racial or gender biases. Additionally, using the stolen data \dash including both the ML model and knowledge of the contents of past application packets \dash the researchers hope to assess the vulnerability of Company~B's AI system to adversarial manipulation. Informed by a scientific understanding of the biases and vulnerabilities in Company~B's AI system, the researchers intend to propose technical and policy mechanisms to mitigate such biases and vulnerabilities in the future.

The researchers know, however, that the data was stolen and shared publicly over the objections of many job applicants. The researchers must choose between doing nothing (not studying the data) or studying the data and reporting on the results. If the researchers study the data and report on their results, they know not to include anything in their publication that could lead to the identification of any of the job applicants. If they study the data, they also know that they must continue to retain a copy of the data even after publishing their results in case their results are challenged, e.g., by Company~B.

\paragraph{On this Scenario}
Adjacent to the computer security research field, the human-computer interaction field has an extensive history of considering the morality of studying data that people might have technically made public but that they might not wish to be used in research or that might cause harms if quoted in a publication, e.g.,~\cite{bruckman2002studying,FieslerTwitter,ZimmerPublic,FieslerReddit,FieslerFan, MarkHam2012}. These works also consider best practices for how to study such data and how to report on the results.

Within the security research community, it is not uncommon to study datasets containing information that users did not intend to be public. A typical example is the study of the contents of stolen password or other databases~\cite{thomas2017ethical}. An adjacent example is the study of anonymized datasets that are, in actuality, not fully anonymized, e.g.,~\cite{netflixprize,adar2007user,sweenymerge}. 
The ubiquity of such studies speaks to at least partial agreement within the community on the morality of such studies in general, though researchers must still pay attention to details. For example, even if researchers study the contents of a leaked password database, they might not include real username and password pairs in a resulting  publications, similar to how human-computer interaction researchers might  not include full quotes in publications even if quoting from public data, e.g., \cite{MarkHam2012,NatureQuote,bruckman2002studying,FieslerTwitter}.

For Scenario~B, we sought a scenario related to stolen data but with content that, by itself, is more sensitive than  usernames and passwords.
We explored numerous possibilities, including (for example) scenarios related to data from victims of intimate partner violence (motivated both by works from within the security community, e.g.,~\cite{chatterjee2018spyware,havron2019clinical},  and earlier works in adjacent areas, e.g.,~\cite{matthews2017stories,freed2018stalker,freed2017digital}), scenarios related to face recognition systems created from scraping ``public'' images (motivated by prior works, e.g.,~\cite{evtimov2021foggysight,shan2020fawkes}), and scenarios related to stolen data about activists during a revolution (motivated by prior works, e.g.,~\cite{daffalla2021defensive}).
Motivated in part by past computer security research on biases and vulnerabilities in remote proctoring software~\cite{burgess2022watching} as well past concerns about biases in job-applicant matching systems, e.g.,~\cite{techreviewAIbias}, we chose to focus on an AI job-applicant matching system: a system for which job applicants might submit an extensive amount of private information. 
We found that the other scenarios were too difficult to fully present and explain in a ``short'' amount of space; the explanation needed to include, for example, a broader context about intimate partner violence or activism.

As with all our scenarios, our goals in Section~\ref{sec:scenarios-initial:criteria} influenced our scenario design. Here we highlight two aspects of this scenario that enable it to meet our ``decisive'' goal.
First, while one might argue that people's right to privacy extends to data that they intended to be private even after others (illegally) made the data public, we make the right to privacy in Scenario~B even more definitive by having those impacted by the data leak explicitly request that all copies of the data be deleted.
Second, if biases are present in the AI system, and if those biases are removed, that would change which applicants are shown to employers. Although preferable in terms of overall fairness, such a change could also do harm, e.g., to the removed applicants. To simplify our consequentialist analyses, we explicitly assume that anyone removed through this process would still be able to find a job that they desire.

\subsection{Scenario C: Inadvertent ``Disclosure''}
\label{sec:scenario:inadvertentdisclosure}

Scenario~C features an ethical dilemma for a conference program committee member. We selected this scenario to be among the three featured because questions of ethics and morality arise not only in research (Scenarios~A and~B), but also during the peer review process (this scenario).
Figure~\ref{fig:scenario:inadvertent_disclosure} in Appendix~\ref{ap:figures} provides a self-contained presentation of this scenario.

\paragraph{Context} A program committee member works for Company C and, as part of the program committee process, encounters a confidential paper submission detailing an undisclosed vulnerability in Company C's product.  Upon reading the  submission, the Company C employee realizes that the vulnerability is very serious and that it will take a significant amount of time to patch. 
The employee feels an obligation to their employer and to Company C's users. But,  the program chairs required all committee members to explicitly agree to maintain the confidentiality of all submissions.
Company C's leadership team decided that the Company C employee should agree to the confidentiality condition and join the program committee. 

\paragraph{The Choice} The employee of Company C must  decide between doing nothing (not disclosing the vulnerability in the paper to Company C) or disclosing the vulnerability to their employer. 

\paragraph{On this Scenario} 
While we are aware of real-world scenarios similar to Scenario~C, we are unaware of written public statements about those situations and consequently include no background citations. Scenario~C is thus based solely on the memories and experiences of the computer security researchers on this team as well as discussions with others. As with our other scenarios, the real world is more complex, with additional options available to the program committee member, e.g., the program committee member could work with the program chairs to determine a course of action. 
And, as one possibility, if the program committee member contacts the program chairs, the program chairs  could assert full decision-making responsibility, thereby not requiring (or even permitting) the Company~C employee to make any subsequent decisions (the Company~C employee already made at least one decision: to discuss with the chairs).

\section{Ethical Frameworks}

\label{sec:frameworks}

Ethical frameworks define approaches for reasoning about whether actions are morally right or wrong. In ethics / moral philosophy, the oft-cited three main ethical frameworks are  consequentialist ethics, deontological ethics, and virtue ethics. A fourth oft-discussed framework is discourse ethics.
We discuss the first two in Section~\ref{sec:frameworks:CandD}. Although our analyses focus on the first two, we discuss the latter two along with several other frameworks, including principlism (featured in the Belmont and Menlo Reports~\cite{Belmont,Menlo}) in  Section~\ref{sec:frameworks:other}.

The frameworks we explore have in some cases evolved over considerable periods of time, with a multitude of contributions, objections, and adaptations. 
There can thus exist a vast variety of different branches and nuances within each framework. Since our goal is to explore moral dilemmas in computer security research from the perspective of different ethical frameworks and not to argue, for example, the benefits of one framework over another or for a new theory of ethics for computer security research, we limit our descriptions to the general features of each framework. Our summaries are sufficient to clearly contrast the different frameworks with each other and to receive clearly distinguishable reasonings and outcomes with regard to our scenarios from Section~\ref{sec:scenarios-initial}. 

While we believe that our summaries are sufficient to enable security researchers to explore their own problems with these frameworks, we defer interested readers to works such as
Anscombe's article ``Modern Moral Philosophy''~\cite{AnscombeJournal}, Baggini and Fosl's book \textit{The Ethics Toolkit}~\cite{EthicsToolkit},
Deigh's book \textit{An Introduction to Ethics}~\cite{DeighIntroEthics},
Driver's book \textit{Ethics:  The Fundamentals}~\cite{DriverEthics}, and Stanford University's online resources~\cite{StanfordEthics} for additional, general information.
For works focused on ethics and technology / engineering, we defer readers to works such as
Floridi's book \textit{The Cambridge Handbook of Information and Computer Ethics}~\cite{FloridiHandbook}, 
Iphofen's book \textit{Handbook of Research Ethics and Scientific Integrity}~\cite{IphofenHandbook}, Quinn's
book \textit{Ethics for the Information Age}~\cite{QuinnEthics}, and Santa Clara University's online resources~\cite{SCU:Ethics}, as well as professional codes of ethics~\cite{NSPE:ethics,ACM:ethics,IEEE:ethics}.
Further, as we stress elsewhere, we are \emph{not} arguing for the application of any of the frameworks we survey; rather, we are arguing for the use of these frameworks as mechanisms to facilitate thoughtful dialog and inquiry while, for example, applying the principles in the Menlo Report~\cite{Menlo}.

\subsection{Consequentialist and Deontological Ethics}
As discussed earlier, we center consequentialist and deontological ethics in our analyses because of their prominence in the field of ethics / moral philosophy and because of their existing role in the computer security research community, e.g., their presence in the Menlo Report~\cite{Menlo}.

\label{sec:frameworks:CandD}

\paragraph{Consequentialist Ethics}
Consequentialism centers the consequences --- the outcomes --- of an action, both positive (benefits) and negative (harms). Each consequentialism comprises of a value theory (e.g., hedonism) and a moral principle (e.g., maximizing overall utility), according to which an action is morally right exactly when there is no other action with better consequences as measured by the respective value 
theory.

Utilitarianism is an example of consequentialism in which positive and negative outcomes are generally assessed with respect to the well-being (welfare) of people. 
We use utilitarianism in the consequentialist analyses in this paper.
Under utilitarianism, the right action is the action that produces the greatest net positive well-being.
There are three main categories of utilitarianism,\footnote{For the purposes of this paper and the decisions in the scenarios, we focus on direct action utilitarianism and ignore rule utilitarianism.} each corresponding to one of three main theories of well-being:
\begin{itemize}
    \item \textbf{Hedonic utilitarianism:} An action is right if it produces the greatest net happiness --- the greatest aggregate happiness over a given set of individuals~\cite{stuart1863utilitarianism}.
    
    \item \textbf{Preference utilitarianism:} An action is right if it enables the greatest number of people to live by their own preferences~\cite{hare1981moral}.

    \item \textbf{Objective list utilitarianism:} An action is right if it produces the greatest net positive impacts on the greatest number of people with respect to an objective list of measures~\cite{parfit1984reasons}; example measures are the levels of one's health, wealth, or access to resources.
\end{itemize}

For objective list utilitarianism, the standards to maximize for are not subjective desires or preferences, but rather ``objective'' (in the sense of ``applicable to all'') measures such as level of health, wealth, and safety (happiness could also be one standard on the list, though some argue that happiness cannot be objectively measured).

These categories are related but distinct. For example, increased health (an objective list measure) can lead to increased happiness (the hedonic measure). Likewise, if someone can live by their preference (preference), then they may be more likely to be happy (hedonic). On the other hand, and as a security-related example, people might prefer to create short passwords or not waste time waiting for software updates 
 to complete (preference), but the use of short passwords and declining software updates could make people's computer systems less secure (an objective list measure).

Rather than rely solely on a single definition of well-being and hence a single category of utilitarianism, those evaluating morality of actions may employ:
\begin{itemize}
    \item \textbf{Pluralistic utilitarianism:} Pluralistic utilitarianism considers happiness, preference,  objective lists, and other forms of benefits / harms in combination.
\end{itemize}

In moral considerations, a central focus is on the question, ``what is the right decision to make?'' However, the question ``did we make the right decision?''\ is equally important, as it deals with questions of (retrospective) responsibility, redress, and retributive justice.
When evaluating the moral quality of an action that has already happened, one view of consequentialism focuses on the actual outcomes regardless of what the likely outcomes were prior to the action.
A probabilistic view of consequentialism asks whether the action was likely to have produced a net positive outcome regardless of whether it actually did so. 
Under the former view, an action that would likely have produced net negative results but that did not is still a right action; under the latter view, the action is not right.

Relatedly, when considering what decision to make, direct action utilitarianism focuses on the outcome of an action. Rule utilitarianism focuses on whether the decision follows rules designed to maximize positive net outcomes. Under rule utilitarianism, an action that causes net harm is still right if it follows rules that, across all scenarios, produce the greatest net positive results.
We designed the scenarios in Section~\ref{sec:scenarios-initial} to highlight key points of consideration about benefits and harms in individual situations and not as vehicles to discuss generalizable rules for the field.
Hence, in Section~\ref{sec:analysis-scenarios-initial}, we adopt a direct action utilitarian perspective.

\paragraph{Deontological Ethics}
Deontological ethics focuses on the moral duties of a given moral actor, such as an individual or an institution. These duties are often specified as direct duties (obligations) against others\footnote{This is one aspect that differentiates deontological reasoning from consequentialism, which at most posits a general duty to be moral (i.e., produce the best outcomes).} \dash i.e., what does one person (morally) owe others~\cite{kant1785grundlegung}? 
These duties are often specified in terms of justice, either as negative duties (refrain from doing harm)\footnote{While consequentialism is also concerned about (overall) harm, it does not hold that there are specific duties to the single individuals not to harm them. Rather, it aggregates harms and benefits, such that one given individual might suffer considerable harm if the net benefit for others is positive.} or as positive duties to certain claims that others have. 
In modern rights-based theories, these duties correspond to (moral) rights of the moral patient to whom the duty is owed. For example, if one person has a right to privacy, others owe this person (the moral patient) a certain behavior associated with that right.\footnote{These rights are often spelled out in Hohfeldian claim rights and corresponding obligations. As in our scenarios, the moral agents (researchers, program committee members, and so on) may incur direct moral obligations, Hohfeld's distinction between claim rights and liberty rights is not important here.}

A defining result of the duty- / rights-based approach of deontological ethics is the focus on the right intention to act. While consequentialist ethics are mainly concerned with the outcomes, deontological ethics ask whether the action is undertaken out of a consideration for one's moral duty, or by some other thought process. Only an action that is performed with the intention to discharge a moral duty is considered moral.\footnote{This is not to say that deontological ethics entirely disregards the outcomes of an action. Neo-Kantian versions like John Rawls' ``difference principle'' (unequal distribution of certain goods is just, as long as it also benefits the least well-off), for example, often add a consequentialist aspect to the otherwise deontological reasoning. However, also in these cases the primary factor is individual rights and the intention to discharge duties associated with these rights.} Kant as one of the main protagonists of deontological ethics distinguishes between acting morally (i.e., out of consideration for one's moral duty) and legally (e.g., out of consideration for an actual legal framework or out of fear of sanctions). For example, completing a human subjects review process, such as an IRB within the U.S., solely because it is a university requirement is not a moral act.

Deontological ethics differ widely in their justification of the respective duties. One historical example is Divine Command and the duty to fulfill God's will.
Most famously, Kantian ethics derives the moral duties from the faculty of reason that human beings have: because we can reason about what to do and thus control our desires, we have the obligation to do so in order to become autonomous (giving ourselves the moral law). And, since we are all potentially autonomous, we have a duty to treat all other human beings as such, i.e., as ``ends and never purely as means'' in Kant's words. A modern version of this Kantian thought is a specific take on contractualism~\cite{scanlon2000we}, which posits that we should act in a way that cannot be reasonably rejected by anyone. 
Natural rights theories, on the other hand, take the idea that human beings as moral actors have certain faculties and justify natural (i.e., unalienable) rights (and corresponding duties) from those faculties for all persons (e.g., John Locke's ``life, liberty, and estate'' or the Virginia Bill of Rights)~\cite{locke1988}.

Given the influence of Kantian deontological ethics on the Belmont Report~\cite{Belmont} and the Principles of Biomedical Ethics~\cite{BeauchampChildress}, and hence on the Menlo Report~\cite{Menlo} and (sometimes implicit) arguments within the computer security research community, we take a Kantian approach to our ethics analyses.
In order to make deontological ethics more tangible for computer security ethics and to contrast it more sharply with consequentialist ethics, in this paper we make a somewhat simplified assumption that deontological ethics conducts moral evaluation in the form of (individual or collective) duties and corresponding (individual) rights, that are spelled out in (absolute) terms of right or wrong. For example, if it is a duty not to harm someone, then killing one person to save five other lives \dash as presented in the classical trolley problem
(Figure~\ref{fig:scenario:trolley}) 
\dash directly interferes with this duty (and the person's right not to be killed) and is therefore wrong, no matter what.\footnote{Of course, there are also pro-tanto-duties, which only hold as long as no more important or pressing duty surfaces. In order to contrast the two traditions more sharply, however, we ignore these in this paper.}

In contrast, consequentialist ethics, as exemplified by utilitarianism, conducts moral evaluation in the form of overall well-being (net utility), which allows comparative evaluations. The state in which only one person is dead  will thus typically be a better state  than the state in which five people are dead. While deontological ethics focuses on the intention (to discharge one's moral duties and to honor the rights of others to be treated in a certain way), consequentialist ethics focuses on the outcome of an action, policy, social practice, and more.

\subsection{Other Ethical Frameworks}
\label{sec:frameworks:other}

While there are other ethical frameworks, for the purpose of this paper we focus on consequentialist and deontological accounts (Section~\ref{sec:frameworks:CandD}).
These frameworks already have a strong presence within the security community, e.g., in the Menlo Report~\cite{Menlo}. Indeed, elements of these frameworks are (at least implicitly) present whenever a researcher weighs benefits and harms or considers human rights. By construction (Section~\ref{sec:scenarios-initial:criteria}), these are also not only the easiest frameworks to apply in the scenarios we present, but also  contrast each other (at least in the simplified scenarios that we have established). 
Nonetheless, we wanted to give a brief overview of some other ethical frameworks that seek to answer the question: What does it mean to act morally? 

\paragraph{Virtue Ethics}
Virtue ethics focuses more on how actors should \emph{be}, i.e., a (morally) virtuous person, than on how they should \emph{act}, since from being a virtuous person, (morally) virtuous actions will follow. In the Western world, the virtue ethical tradition can be traced back to Plato and Aristotle, who argued that people should cultivate and internalize virtuous moral character traits. Once internalized, a virtuous person would act virtuously \dash would choose the right action \dash by habit and by instinct. While ``by habit and by instinct'' at least for some authors necessitates immediacy, virtuous persons will still often use practical wisdom (``phronesis'') in order to assess new situations and how they relate to internalized virtues. As this practical wisdom is itself a virtue, employing it in the right fashion and right instances also shows a virtuous person.
While (in this Aristotelian sense) truly virtuous persons act on their internalized moral dispositions, they may also take time to thoughtfully consider the available actions, the expected outcomes, their responsibilities, and more, before deciding upon which action to take. Elements of virtue ethics manifests in the security community,
for example, through the internalization of approaches to ethics after repeatedly considering ethics and security research over time, or whenever a program committee asks the question,
``did the authors realize that they might have caused harm?''

While much of ethics / moral philosophy centers  Western history,  the internalization of virtuous character traits is present in numerous other traditions around the world, such as the \textit{yamas} (external ethical practices) and \textit{niyamas} (internal ethical practices) of the eight-fold Yogic traditions and the striving for \textit{mushin} (an empty mind without motives or ego) in Japanese tradition. Likewise, and relatedly, Buddhist ethics and Confucian ethics also center virtues~\cite{puett2016path}.

\paragraph{Discourse Ethics}
Unlike consequentialist, deontological, and virtue ethics, which focus on the actions or actors, discourse ethics center a \emph{process} for moral decision making, namely an \emph{idealized moral discourse} that is egalitarian, inclusive, and principally interminable. Under discourse ethics, the belief is that whenever this idealized moral discourse nears a consensus, then this consensus is as close as we will ever get to moral truth. Discourse ethics provides a process for determining the morally right action even when, \textit{a priori}, there is  disagreement in what is morally right.
In order to make this framework operationalizable, a key consideration has to be how to transform this justificatory claim about moral truth in a principally boundless moral community into the moral status of a real discourse within the computer security community. Another question is what constitutes ``idealized moral discourse' within the realm of computer security. For example, could discussions within ethics review committees be regarded as an \dash albeit very specific and expert \dash idealized moral discourse?

\paragraph{Principlism Ethics}
Principlism ethics believes that there are relatively uncontroversial principles upon which most moral theories converge, i.e., that there are a few principles central to the other frameworks combined. These principles should be used as a starting point to assess the morality of actions, institutions, policies, and more. One example is the four principles of biomedical ethics from Beauchamp and Childress~\cite{BeauchampChildress}: 
respect for autonomy, 
beneficence,
non-maleficence, and 
justice.
Also published in the same era (1970s) are the principles in the Belmont Report~\cite{Belmont}, which focuses on the protection of human subjects in research. The Belmont principles are: respect for persons, beneficence, and justice.

The Belmont~\cite{Belmont} and the Beauchamp and Childress~\cite{BeauchampChildress} principles derive from both consequentialist and deontological ethics. Under principlism ethics, when faced with a moral decision, the actors should evaluate their courses of action with respect to these identified, universal principles. When conflicts arise between the application of different principles, e.g., between a consequentialist-derived principle and a deontological-derived principle, a process must be used to resolve those conflicts.
The Menlo Report articulates the application of the Belmont Report's principles to the computing research field~\cite{Menlo}. To make  use of principlism for computer security scenarios, one must specify these principles at a concrete level and also formulate rules of priority in cases of value conflicts.

\paragraph{Emancipatory Ethics} 
Emancipatory ethics is the kind of ethics that Critical Theory would call its own, if Critical Theory entertained a dedicated ethics.\footnote{The reluctance of many proponents of Critical Theory to talk about ethics or ``an'' ethics is often due to the perceived function of ethics to ``tell others what is right or wrong''. This goes directly against the emancipatory endeavor of Critical Theory to further ``true'' (ethical) self-reflection~\cite{horkheimer1937traditional}.}  
This kind of ethics is our umbrella term for more specific ethical enterprises, such as Ideology Critique, or certain versions of Care Ethics and Feminist Ethics. Common to all of these is the focus on the self-emancipation of individuals or groups that are in some way oppressed or marginalized. In order to achieve this outcome, emancipatory ethics centers the emancipated life \dash a life in which everyone can know and pursue their own ``true interests''~\cite{geuss1981idea}.  
Emancipatory ethics therefore does not provide a framework to distinguish (moral) right from wrong, and not even an account of the conditions necessary such that those affected may be able to do so themselves. Rather, it criticizes current conditions as detrimental to a self-reflection on the affected's ``true interest'' and the individual and social transformations necessary to realize these interests~\cite{honneth2014freedom}. 
More concretely, it analyzes social structures and relations in order to uncover and understand inequities, power structures, oppression, and discrimination, which will prevent those affected by these structures to ``truly'' self-reflect on their interests and associated moral obligations and to express them freely~\cite{horkheimer1947theodor}.  
Under emancipatory ethics, there is no right moral action. Rather, it points to inequities, power structures, oppression, and more, not because of them being morally wrong, but because they inhibit the reflective moral process of those affected by those structures.
Moreover, the belief is that those in power are often as tied by these structures as the oppressed, but with more to lose. Hence, under emancipatory ethics, it is essential to (1) highlight the results of the above-mentioned analyses and (2) thereby enable those subjected to these power asymmetries to contest them and demand change. An example of (2) would be including those with less power in the process of determining what is right, or inviting them to lead that process. As another example of the application of emancipatory ethics, if consequentialist- or deontological-like analyses are performed,  then inequities, power structures, oppression, and discrimination must be considered centrally.

\section{Analysis of Scenarios A, B, and C}

\label{sec:analysis-scenarios-initial}

We now turn to using consequentialist and deontological ethics (Section~\ref{sec:frameworks}) to analyze the scenarios in
Section~\ref{sec:scenarios-initial}.
We encourage readers to review Scenarios~A, B, and~C first and consider what decisions they would make before reading our analyses. If readers wish, they may complete an online Google Form with their decisions (link available at \url{https://securityethics.cs.washington.edu}) and, upon doing so, see how others chose to respond.\footnote{
The Google Form is anonymous \dash it requires Google authentication but does not reveal any identifiers to the authors of this paper. When interpreting the results of this form, we caution that no mechanisms, other than Google authentication, are used to protect against the use of different Google accounts to vote multiple times.}

As we noted in the discussion of our scenario criteria (Section~\ref{sec:scenarios-initial:criteria}), we designed our scenarios to reflect the real world but to also facilitate clear, precise analyses.
Real-world scenarios can be and often are significantly more complicated and may present more than a binary option to the decision-makers. 
Given the potential for greater uncertainty in real-world scenarios, decisions may require
an ethical risk assessment about uncertain future states. Further, decision-makers must also consider the law, in addition to ethics.

\subsection{Analysis of Scenario A (Medical Device Vulnerability)}
\label{sec:analysis-scenarios-initial:medical}

Here we consider the medical device vulnerability scenario from Section~\ref{sec:scenario:medical} (see also Figure~\ref{fig:scenario:medical_device} in Appendix~\ref{ap:figures}).

\paragraph{Consequentialist Ethics}
Physical health in this scenario is an objective measure; if a patient chooses to remove a device or chooses not to obtain one because of a known vulnerability, then they would have a shorter life expectancy.
Psychological health in this scenario is also an objective measure; if a patient knows about the vulnerability and still chooses to keep or get the implant, then they could live in fear of a security incident even though the likelihood of an incident is zero (by scenario construction).\footnote{In the real world, decision-makers must also consider family members, loved ones, and other stakeholders; we focus on patients for expositional simplicity.}

From a hedonic perspective, the knowledge that one has a shorter life expectancy (if they do not have the device) or the fear of a security incident (if they have the device) could lead to decreased happiness. In addition, the fact that removing or not opting for the device will result in ten years less of potential happiness may also significantly decrease overall happiness. 
From a preference utilitarian perspective, under the assumption 
that most patients would prefer not to  learn about the vulnerability, then not disclosing the vulnerability would maximize the ability of people to live by their preference.

Hence, the morally correct decision is to
\emph{not disclose the vulnerability}.

\paragraph{Deontological Ethics}
Under deontological ethics, the researchers have a duty to respect people's right to informed consent and the right to self-agency. In the medical context, this right to informed consent manifests (for example) as warnings in TV advertisements for medicines. These are fundamental human rights, and not disclosing the vulnerability would violate those rights.
Hence, the morally correct decision is to \textit{disclose the vulnerability}. This conclusion is correct even if most people would have preferred not to know about the vulnerability.

\paragraph{Informed by the Real World, Not Real}
As discussed in Section~\ref{sec:scenarios-initial}, although real-world observations and experiences  informed our scenario designs, there are gaps between our scenarios and what one might encounter in the real-world.
Rather than choose from one of only two provided (binary) options, the researchers might, for example, choose to involve others in the decision-making process or cede the decision responsibility to another entity entirely. In the U.S., the FDA \dash not the researchers \dash could make or strongly contribute to the decision on whether to disclose the vulnerability to the public. Should they choose to disclose the vulnerability, they might work with healthcare providers to thoughtfully and conscientiously craft the message, thereby reducing patient alarm. 
Given the medical and security contexts, the decision-makers might leverage the Principles of Biomedical Ethics~\cite{BeauchampChildress} and the Menlo Report~\cite{Menlo}. Thus, even if the decision-makers do not solely rely on consequentialist or deontological analyses, and indeed consequentialist and deontological ethics both have limitations, consequentialist and deontological thinking may be part of the final decision-making process.

In Scenario~A, we made the assumption that there is \emph{zero} likelihood of the vulnerability being exploited regardless of whether or not the vulnerability is made public. We could have instead provided probability distributions for the likelihood of exploitation both if the vulnerability is made public and if it is not and then, for our consequentialist analysis, we could have calculated the likely overall benefits and harms for each decision. From a deontological perspective, if the public vulnerability disclosure would result in \emph{more} people's devices being compromised than would be the case if there was not a public disclosure, then we would need to consider the negative impact of the public disclosure on those people's rights.
By fixing the exploit probability at zero, our work is able to focus on comparing and contrasting the different ethical traditions rather than providing lengthy empirical analyses.

The above discussion points to another challenge with ethical decision making in the real world: uncertainty.
In the real-world, a decision-maker might encounter questions that they cannot precisely answer, such as: Do all potential adversaries already know about the vulnerability? If not, then the public disclosure of the vulnerability might increase the exploit probability. On the other hand, if so, then the public disclosure of the vulnerability might not increase the exploit probability. That is unless, for example, the public disclosure of the vulnerability results in adversaries being more comfortable using their knowledge of the vulnerability. Would they be more comfortable?
Or, supposing that adversaries do not already know about the vulnerability, what is the likelihood that they might discover the vulnerability themselves?

\subsection{Analysis of Scenario B (Studying Immorally Obtained Data)}
\label{sec:analysis-scenarios-initial:immoraldata}

We now turn to analyzing the scenario in Section~\ref{sec:scenario:immoraldata}  (see also Figure~\ref{fig:scenario:immoral_data_jobs} in Appendix~\ref{ap:figures}).

\paragraph{Consequentialist Ethics}
    Being able to find a job that one is qualified for is an objective measure of well-being in this scenario. The research has the potential to uncover biases or attack capabilities that can limit people's ability to find jobs. By proposing mechanisms to mitigate these biases or vulnerabilities, the research output can improve the ability of people to find such jobs.\footnote{Recall from Section~\ref{sec:scenario:immoraldata} that addressing biases will improve the ability of some people to find a job (those impacted by biases) but, to simplify the analysis, will not negatively impact the ability of other people to find a job.}
    Thus, from an objective list utilitarian perspective, the benefits of studying the data is high. Further, the data is already ``public'' and hence harm to job applicants has already happened.
    Further, the number of people harmed by the theft and release of the data is comparatively small compared to the one hundred-fold prediction of future use.
    Thus, from an objective list utilitarian perspective, the morally correct decision is \emph{to study} the data.
    
    A hedonic or preference utilitarianist would, respectively, observe that analyzing the data could degrade the happiness of the people whose data was stolen and would also prevent them to live by their preference, if they would prefer that the data not be studied. However, the number of people who would benefit (in both happiness and the ability to live by their own preferences) after the data is studied is far greater. Hence, even with the hedonic and preference utilitarianist frameworks, the morally correct decision is \emph{to study} the data. 
    
\paragraph{Deontological Ethics}
Taking a Kantian deontological view, we observe that people have inalienable rights, including agency and privacy. Those rights extend to data intended to be private, whether it is private or not. Further, even if the right to privacy did not extend to adversarially-released data after it becomes public, in this scenario, the victims of the data breach have explicitly requested that their data be deleted everywhere.
In order to do their research, the researchers would need to retain a copy of the data, thereby disrespecting the request to delete all data copies. They would also need to retain a copy \emph{after} their research is complete in case their results are challenged, e.g., by Company~B. 

One might observe that future job applicants have a right to be treated fairly during the job application process, that the research results could result in a more fair AI system, and hence ultimately that the research would result in greater respect for the rights of future job applicants. However, under Kantian deontological ethics, individuals enjoy dignity. In Kant's own terms, this means that individuals may \emph{not only} be treated as a means, but also always as \emph{an end in itself}. Violating privacy rights in order to study the data (and prevent future harm) amounts to treating those whose data is studied solely as means to a different end, and is therefore  wrong.

One might ask whether it would be appropriate to contact victims of the data breach and ask if their data can be retained and used for research \dash i.e., to obtain those victims' informed consent. This scenario does not present that option to the researchers. However, even if it did, under a deontological perspective, the act of asking a victim for informed consent in this scenario requires using the stolen data (to obtain victim identity or contact information); using the stolen data in this way is already a violation of privacy. Further, the act of contacting the victims could have unknown consequences.

Therefore, under Kantian deontological ethics, the stolen data should \emph{not be studied}.

\paragraph{Informed by the Real World, Not Real} As with Scenario~A, this scenario is informed by real-world experiences and observations but is not real. Researchers in the real world might have a mandatory first step prior to analyzing the data, e.g., if the researchers are in the U.S., they should work with their institution's IRB. The IRB would leverage the principles in the Belmont Report~\cite{Belmont}, which itself includes both consequentialist and deontological reasoning. The researchers may also seek input from others. For example, they may seek input from AI and security ethics experts, who might then also reference consequentialist, deontological, and other ethical frameworks. The researchers might also seek input from populations impacted by the study or non-study of the data, including representatives of people impacted by the data breach and representatives of people who could be harmed by the perpetuation of biases in Company~B's AI system. Moreover, the researchers might offer these groups the option to lead the decision on whether to study the data.

\subsection{Analysis of Scenario C (Inadvertent Data ``Disclosure'')}
\label{sec:analysis-scenarios-initial:inadvertentdisclosure}

We now turn to  the scenario in Section~\ref{sec:scenario:inadvertentdisclosure} (see also Figure~\ref{fig:scenario:inadvertent_disclosure} in Appendix~\ref{ap:figures}).

\paragraph{Consequentialist Ethics}
    The consequentialist must weight harms against benefits. There are harms to authors if the employee of Company C discloses the vulnerability to Company~C \dash the authors will not be able to disclose at their preferred time (perference utilitarianism) and may be unhappy (hedonic utilitarianism) and may have their careers or other aspects of their lives negatively impacted if the early disclosure to Company C limits their impact or ability to publish (career advancement could be a measure of well-being per objective list utilitarianism).
    
    However, the harms to Company C's users if the employee does not disclose the vulnerability to Company C is much greater \dash without early disclosure, Company C will not be able to protect their users and, as a result, millions of people around the world could be significantly harmed. 
    
    Hence, the morally correct action is for the employee
    \emph{to disclose} the vulnerability internally to Company C.

\paragraph{Deontological Ethics}
    The employee of Company C may feel a sense of duty to their company and to their company's users. However, program committee members also have a duty to respect the autonomy of authors and a duty to respect the confidentiality of the peer review process. Moreover, the employee of Company C agreed to respect this duty when they joined the program committee, as did Company C's leadership team when they granted the employee permission to join the program committee. Moreover, from a Kantian reasoning, the employee could not form a maxim that allowed breaking the confidentiality promise, as otherwise the peer review process and the institution of program committees would not be possible. 
 
    Thus, the morally correct thing for the Company C employee to do is respect the rights of the authors and the confidentiality of the review process and \emph{not disclose the vulnerability} to Company C.

\paragraph{Informed by the Real World, Not Real} In a real-world scenario, the employee of Company~C might not make the decision on their own. For example, rather than decide between the two options we presented, they might first reach out to the program chairs and ask them to give advice or render a decision.
The program chairs might then explore questions such as: should they reach out to the paper's authors, asking for more information about the disclosure timeline? If  Company~C's employee assesses the harms of  the vulnerability as significant, should the program chairs ask the authors to disclose to Company~C right away? What are the impacts on the scientific peer review process if the program chairs ask the authors to disclose to Company~C? Would the authors feel compelled to grant permission because they want their paper to be accepted even if granting permission is not in their best interest? Since not all companies have members on the program committee, is it morally right to give this company (and their users) advance notice of a vulnerability (even with author permission) solely because an employee of their company is on the program committee? 

\section{Additional Scenario Contributions}
\label{sec:scenarios-more}

Although Scenarios~A, B, and~C, along with their  analyses, are our core contributions, in developing our initial sets of scenarios (per Section~\ref{sec:scenarios-initial:criteria}), we identified numerous other scenarios that we believed would be valuable to document and make available for community discussion. 

We summarize several additional scenarios, and the reasons for their creation, here. Appendix~\ref{ap:scenarios-more-full} provides more detailed descriptions of these scenarios. Reading Appendix~\ref{ap:scenarios-more-full} is not necessary to understand the core contributions of this paper. Still, the contents of the appendix may be of interest to those wishing to dive even more deeply into an exploration of computer security research scenarios from the perspective of ethics /  moral philosophy. Scenarios~D$^*$ might be of particular interest to researchers working on vulnerability disclosures and Scenarios~E$^*$ and~F might be of particular interest to program committee members.
As with the Google Form for Scenarios~A, B, and ~C, we provide Google Forms for Scenarios~D$^*$, E$^*$, and~F at \url{https://securityethics.cs.washington.edu}.

\paragraph{Scenarios~D$^*$: Vulnerability Disclosure}
\label{sec:scenarios:more:disclosure}
This is a family of scenarios, D1 through D7, that feature different considerations with respect to vulnerability disclosure. In the base scenario, Scenario~D1, the morally right decision is obvious across consequentialist and deontological ethics: researchers should disclose a vulnerability to a manufacturer first, before disclosing the vulnerability to the public. (As with all our scenarios, we intentionally simplify our scenarios and limit options; in the real-world, other options might include anonymous disclosure or disclosure through an entity such as CERT.)

The remaining scenarios in this sequence provide variations that increase the complexity of the moral decision. For example, what if the company is litigious and would block the publication of the research after receiving a disclosure (Scenario~D2)? To aid in the analysis, in Scenario~D2, we also assume that the company does not care about security and will not work on a patch even after a private vulnerability disclosure, and hence disclosing privately to the company first would ultimately lead to the greatest harm to users (the researchers cannot publicly discuss their results, and the company will leave users vulnerable).

Next, what if the company in question is in an industry that is highly litigious but it is unknown whether the company itself is litigious (Scenario~D3)? And, for Scenario~D3, it is also known that the company is in an industry that, as a whole, does not care about security and will not begin developing a patch even after a private disclosure, but it is also unknown whether the company itself cares about security or not.

Next, what if there is significant uncertainty on the likelihood of adversaries to discover the vulnerability on their own (Scenario~D4)? Or, what if the company is litigious but now it is known that the company cares significantly about security: it would immediately begin working on a patch while at the same time entangling the researchers in a lawsuit (Scenario~D5)?

Next, does the calculus change if the lead researcher is a PhD student and the research is the final piece of their dissertation? If the research gets entangled in a legal battle, the PhD student cannot graduate and must either decline the industry offer that they already accepted and stay in graduate school longer or leave graduate school without a PhD. How should the PhD advisor handle such a situation (Scenario~D6)?

Lastly, does the calculus change if the company's users are, for the most part, mostly engaged in an illegal activity, e.g., the vulnerability, if exploited, would allow someone to learn the names and email addresses of people sharing non-consensual explicit material (so-called revenge porn) (Scenario~D7)?

\paragraph{Scenarios E$^*$: Submission Raises Ethical Concerns}
\label{sec:scenarios:more:submission}
This family of scenarios are all related to the following situation: a program committee reviews a paper that reports on research that should not have been done, e.g., an Internet crawl that caused  insulin pump machines in hospitals to crash. The paper is long, and the part that should not have been done (the Internet crawl) is confined to only one small section of the paper (Section 9.3). What decision should the program committee make about the paper?

In the base case (Scenario~E1), the program committee has two options: reject the paper, or accept it without any required modification; if the latter option is selected, the authors will receive reviews from the program committee and will have the option to voluntarily revise the paper as they see fit. While offering only two options is consistent with our ``binary decisions'' criteria goal in Section~\ref{sec:scenarios-initial:criteria}, we decided to add an additional option for Scenario~E2: accept the paper with the  relevant results (Section 9.3) removed. We add another additional option for Scenario~E3: accept the paper but attach a note (written by the program committee) to the paper that explains the ethical concerns. We added these additional options because our goals with these scenarios are different than in Section~\ref{sec:scenarios-initial:criteria}. A central goal here is to offer scenarios that facilitate conversations within the community, and program committees will likely not consider only the initial two binary options.

Our next scenario asks whether the program committee would make a different decision if they know, for certain, that the researchers did extensive testing and tried to minimize the likeilhood of crashes and, in fact,  thought that they had done so (Scenario~E4). And, does the calculus of the program committee change if they learn that the authors \emph{did} know about the potential for crashes (in general, not for insulin pumps), but that the authors felt the moral responsibility to do their crawls anyway because the crawling results will benefit society (Scenario~E5)?

Or, what if the researchers knew there was a risk of crashes (in general, not for insulin pumps) but decided to proceed anyway because they thought that the results would increase the likelihood of their paper being published (Scenario~E6)? Or, what if the scenario is exactly like Scenario~E6 but the researchers were simply lucky and no crashes happened \dash does the absence of any actual harms, even though the researchers believed that their crawls could cause crashes, change the program committee's calculus (Scenario~E7)? Or, what if the scenario is exactly like Scenario~E6 except that the program committee strongly believes that the results in Section 9.3 of the submission should be public, and that not publishing the findings in Section 9.3 of the submission would result in harms to people (Scenario~E8)? Or, does the program committee's calculus change if the researchers were previously naive and, after learning about the impact of their work on insulin pumps, the researchers express significant regret (Scenario~E9)?

\paragraph{Scenario~F: Response to Submission Rejection} 
\label{sec:scenarios:more:rejection}
Scenario~F is a continuation of Scenario~E1, from the perspective of the researchers. Suppose the researchers receive a rejection, along with an explanation from the program committee about how the crawls in Section 9.3 of the submission caused insulin pumps to crash. What should the researchers do? Should they stop working on the project? Should they submit the paper, unmodified, to a new conference? Should they remove Section 9.3 of their paper, pretend like the crawls never existed, and submit to a new conference? Should they add a note to Section 9.3 that explains the harms that their crawls had and then submit to a new conference?

\section{Discussion}
\label{sec:discussion}

\subsection{Reflection on Analyses}

We begin by reflecting upon our analyses and summarizing key points and observations. 
We are not claiming that all these reflections and observations are novel and, indeed, many ideas herein are likely familiar to those with expertise in ethics; see Section~\ref{sec:frameworks} for a survey of resources on ethics / moral philosophy. We offer these reflections and observations because we hope that they
can serve to further our community's collective thoughts and perspectives on ethics and computer security.

\paragraph{Different Frameworks Can Lead to Different Conclusions}
For some moral questions, different ethical frameworks lead to \textit{different} conclusions regarding what is right and wrong. 

\paragraph{Different Frameworks Can Lead to the Same Conclusion}
For other moral questions, different ethical frameworks lead to the \textit{same} conclusion regarding what is right and wrong.

\paragraph{A Framework Can Fail to Reach a Conclusion} We intentionally designed our scenarios to be ``decisive'', per the goals in Section~\ref{sec:scenarios-initial:criteria}; real-world scenarios may \emph{not} be decisive and may \emph{not} lead to conclusive decisions under either the consequentialist or deontological frameworks. Also, it could be the case that under a framework a certain action is morally permitted, i.e., not necessarily required but also not forbidden.

\paragraph{Ethical Frameworks Can Provide Tools for Discussion} 
What
should one do when there are differences of opinion or lack of clarity into what constitutes the right decision? Here is where the tools \dash the frameworks \dash from ethics / moral philosophy can help. In short, they can help decision-makers thoughtfully, methodically, and articulately analyze  moral questions.

In discussions of right or wrong, when there is disagreement, we suggest first surfacing the communicants' underlying values and their  frameworks of consideration. Simply knowing that another communicant is centering different values and a different framework may help further a collaborative discussion.

\paragraph{Ethical Frameworks Can Provide Tools for Thought}
In this work, we primarily consider consequentialist and deontological ethics.
Both of these frameworks have limitations, and we are \emph{not} advocating for strict adherence to either of them. 
In fact, it is not uncommon for people \dash including modern ethicists \dash to include elements of  multiple frameworks (consequentialist, deontological, and other) as they reason through decisions. Within the security research community, the Menlo Report~\cite{Menlo} includes both consequentialist and deontological elements, for example.

On the one hand, the  observation above might call into question the value of articulating ethical frameworks in the first place: if people are not strictly consequentialist or deontological, what value is there in exploring scenarios from strict consequentialist or deontological perspectives? We argue that precise analyses of scenarios under different perspectives can help the decision-maker in multiple ways.
At a minimum, precise thinking via the ethical frameworks
can help slow the decision-making process and encourage thoughtful reflection and contemplation. Additionally, the frameworks can help decision-makers
identify which parts of arguments they agree with and which parts they do not and, by doing so, help the decision-maker better articulate their own arguments, even if their arguments are neither consequentialist nor deontological.

\paragraph{Sometimes the Morally Correct Action is Not in the Best Interest of the Decision-Maker} In Scenarios~A, B, and~C, we tried to minimize the impact of either decision on the decision-makers themselves. Thus, the decision-makers could focus on the impacts on and rights of others. In the real world, a decision-maker's decision might also impact themselves (e.g., a researcher might desire a publication, and the decision on how to proceed might impact their ability to publish). We explore such situations in 
Scenarios~D$^*$ and~F. 
In short, sometimes the morally right decision might \emph{not} be the decision that seems to be in the best interest of the decision-maker.

\paragraph{Shifting Morality Earlier} Our scenarios all feature moral questions for decision-makers. However, one might ask (not just for our scenarios, but for the field) how to shift questions of morality earlier, such that the scenarios we consider (or the real world encounters) do not come up. In a trolley problem, an example of ``shifting earlier'' might be to ensure that all trolleys have better, more resilient brakes. For Scenario~A, code escrow might enable the patching of devices even after the manufacturer ceases operation. For Scenario~B, the researchers would not have needed to study the data if the underlying AI algorithms were already unbiased and secure. For Scenario~C, the situation could be mitigated if the conference had pre-specified rules for such situations or if the conference required disclosure before submission; of course, whether such a rule should be in place raises its own ethical questions, as exhibited (for example) by the role of legal threats in 
Scenarios~D$^*$.

\paragraph{On Uncertainty}
Within the computer security field, there are significant elements of uncertainty. One challenging element of uncertainty surrounds that of the adversary. It is often not known to decision makers when or if adversaries might manifest. Further, precise adversarial capabilities are seldom known to decision makers in advance of their manifestation. Additionally, there is generally uncertainty regarding what unknown vulnerabilities a system might have. For our core scenarios, we aimed to reduce uncertainty through the concrete, precise description of outcomes for each decision option.
A challenge for the computer security research community, and an opportunity for ethics and moral philosophy researchers, is to formulate frameworks and candidate approaches for ethical decision making in the presence of uncertainty, including uncertainty about adversaries, adversarial actions, and unknown vulnerabilities.
(It is because of the challenges imposed by uncertainties that we explicitly incorporate uncertainty into 
Scenarios~D$^*$.)

\paragraph{On the Role of Details and Time}
Ethical assessments not only depend on moral arguments (e.g., whether we should maximize utility understood as $X$, whether one has a right to $Y$ or a duty towards person $Z$), but also on non-moral considerations about the specifics of a given context. Therefore, it is difficult to provide ethically sound judgments that persist over time (as the field of computer security evolves and learns more), as well as across the nuances of different scenarios.
On the latter, we refer to the D$^*$~and~E$^*$ scenario sequences  
and the impacts of scenario changes on what one might perceive as morally right or wrong.
Likewise, one might consider the impact of changes to the specifics of Scenarios~A, B, and~C, e.g., if impacted job applicants did not explicitly request the deletion of stolen data in Scenario~B or if the program chairs did not explicitly mandate confidentiality in Scenario~C.
Further, our field's understanding of harms and adverse impacts evolve over time, e.g., as the field's knowledge of adversarial capabilities or the (possibly) harmful consequences of a research method matures.
Hence, what might be seen as morally right or permissible at one point in time may, at a later date, be seen as neither morally right nor permissible, and vice versa. Collectively, these are \textit{not} reasons not to strive for detailed, specific rules (or at least guidelines) for ethically sound behavior. Rather, we suggest that the specific contexts are among the factors that the security community must consider if it were to do so. 
We argue that the security community should therefore formulate the rules as specific as possible \textit{and} as general as necessary in order to abstract from the details of a given context. A first consideration here could be that (in much the same way as in legal texts) the impartial nature of morality requires the treatment of like cases alike. Therefore, guidelines should address types of cases, rather than single out individual cases, but with consideration both for when and where generalizability applies and when there may be challenges to generalization.

\subsection{For Consideration}

With the background of our results and our reflections, we now present a collection of considerations for members of the security research community.

\paragraph{For Decision-Makers} Decision-makers (researchers, program committees, others) should consider ethics \emph{before} making decisions, rather than after. For certain moral dilemmas (e.g., Scenarios~A, B, and~C), it is possible to pick an outcome and then find the ethical framework that justifies that outcome. We do \emph{not} argue for this practice. Instead, decision-makers should let the decision follow from a disinterested ethical analysis. Toward facilitating disinterested analyses, we encourage decision-makers to explicitly enumerate and articulate any interests that they might have in the results of the decision; such an articulation could be included as part of a positionality statement in a paper.

Additionally, when discussing Scenarios~A, B, and~C in Section~\ref{sec:analysis-scenarios-initial}, we often write that in the real world there may be more collaborative efforts to decide the right way forward. This is certainly the case, but also in these collaborative efforts decision making has to take place, and moral decision making would follow the same (consequentialist or deontological) reasons. 
One could imagine that all persons involved form their own ethical judgments and then discuss amongst themselves, using the very arguments that we use in this work. Or, they could discuss to form individual moral judgments, and consider the very arguments that we use here collectively.

\paragraph{For Researchers Writing Papers} For researchers new to ethics, the Menlo Report~\cite{Menlo} provides concrete guidance. The Menlo Report and other ethical frameworks can help researchers reach a conclusion about what is morally right.

This paper, we hope, can help researchers consider and  discuss morality when there are differences of opinion or uncertainty regarding what to do.
Because (1) we believe that the field can grow through the explicit articulation of ethical thought and (2) there can be differences in ethical perspectives and thought (as Section~\ref{sec:analysis-scenarios-initial} shows), we encourage researchers to do more than just apply a single approach (consequentialist, deontological, the principles in the Menlo Report, or otherwise) and then act accordingly. Rather, we encourage researchers to conduct analyses under multiple ethical frameworks \textit{and} include the reasoning for their decisions under the multiple frameworks in their paper submissions and publications. If the frameworks lead to the \textit{same} conclusion, the inclusion of multiple arguments can strengthen the paper's ethics section and can serve as part of the growing foundation for ethical thought in the field. If different frameworks lead to \textit{different} conclusions, and the authors proceed with what is considered morally right under one framework but morally wrong under another, then surfacing those different considerations and the final thought processes can be particularly valuable. For example, if papers start including analyses under multiple frameworks, then such analyses could become the norm and published analyses could become additional guides for future researchers.

To aid in the above, 
we propose a process that we call \textit{ethics modeling}. 
This process builds on the Menlo Report~\cite{Menlo} and other approaches for evaluating ethics~\cite{manzeschke2015meestar} 
as well as on some approaches to threat modeling. 
Namely, we suggest that researchers first do a stakeholder analysis to identify all stakeholders potentially impacted by the decision, e.g., using methods from value sensitive design~\cite{friedman2019value}. Then, for each stakeholder, we suggest  explicitly identifying the assets that might be impacted by the possible decisions. Then, for each possible decision, for each stakeholder, and for each asset, enumerate the benefits / harms (consequentialist ethics) and the rights supported / violated (deontological ethics).
The benefits / harms and rights analyses should consider situations in which no adversaries manifest and situations in which adversaries manifest. 
We call this process as \textit{ethics modeling} because it combines elements of both ethical analyses and threat modeling.

We further encourage researchers to become familiar with ethical frameworks not deeply considered in this work. An example in the context of computer security and victims of intimate partner violence is care ethics, as considered in Section 6.2 of Tseng et al.~\cite{EmilyCare}.

\paragraph{For Program Committees Discussing Submissions} We encourage program committees and paper reviewers to become familiar with the different ethical frameworks. When questions of ethics arise in the review process, we encourage program committee discussions to explicitly reference not just \emph{what} the discussants believe is morally right and wrong, but \emph{why} they believe that. The latter \dash the why \dash can explicitly refer to analyses under one or more ethical frameworks. 
Further, 
we encourage reviewers to strive to infer what ethical framework or approaches the authors took, if any and if not explicitly articulated, and to consider that the authors may have centered a framework or approach that differs from that of (at least some of) the reviewers. The authors' approach might have been shaped by an academic environment or culture different from the reviewers' own, for example.

\paragraph{For the Community} We encourage the community at large to familiarize themselves with different ethical frameworks. Those community discussions could leverage the scenarios that we developed over the course of this research. For example, prior to reviewing papers, a program committee could, together, discuss the committee's perspective on the right decisions for the scenarios that we present.
From preliminary conversations with members of our community, we believe that such discussions will not lead to a universal consensus. But we believe that the resulting conversations, and the points raised, would be helpful for those community members as they, for example, embark on reviewing papers with possible ethical concerns. 

Additionally, we encourage continued community-wide conversations around 
infrastructure support for proactive, pre-reseach considerations of ethics and morality beyond what is traditionally covered by IRB.
For example, the security community might draw inspiration from the Ethics and Society Review Board as implemented by Stanford HAI~\cite{BernsteinESR} as well as existing approaches for peer-review prior to the implementation of a research method, e.g.,~\cite{COSregisteredreports}. As with research efforts and program committee reviews, we believe that such evaluations would benefit from considerations under, or at least awareness of, multiple ethical frameworks.

\paragraph{For Educators} We encourage educators to include explicit discussions of ethics and ethical frameworks in their courses if they are not already doing so.
Our Scenarios~A, B, and~C, by design, do not lead to obvious right and wrong answers. As a result, we have found that our scenarios are particularly conducive to conversations in classes. Educators are welcome to use our scenarios in their classes as well. A companion slide deck is available at \url{https://securityethics.cs.washington.edu}.
If educators wish to create new scenarios, we encourage them to consider scenarios that meet the design criteria in Section~\ref{sec:scenarios-initial:criteria}.

\paragraph{For Industry and Government}
Although we have scoped our work to focus on computer security research, we believe that this work may also be of interest to those in industry, government, and other sectors. One concrete suggestion, as articulated by an anonymous reviewer and included with permission, is for the Internet Engineering Task Force (IETF) to consider requiring an ``Ethical Considerations under Multiple Frameworks'' section in each Internet-Draft, much like the present requirement of a ``Security Considerations'' section.

\paragraph{For Everyone} Creating ethical norms for computer security research is fundamentally challenging because different ethical frameworks can lead to different conclusions about right and wrong. We believe that a more  achievable near-term goal is the creation of extensive sets of case studies (like our scenarios) that community members can discuss and learn from.

\paragraph{For Us} 
Although we are confident that our dilemmas in Scenarios~A, B, and~C are true dilemmas (per our criteria in Section~\ref{sec:scenarios-initial:criteria}, our validation methodology, and extensive iteration and discussion), this version of our paper does not report concrete data (as doing so is not the goal of this paper). Our ongoing work seeks to provide such concrete data across cultures and communities.  We are additionally preparing other computer scenario descriptions, in the format of the scenarios in this paper, for community consideration. 
As we create these scenarios, we will add them to \url{https://securityethics.cs.washington.edu/}.

\section{Conclusions}

In this paper, we embark on a research collaboration spanning (1) ethics / moral philosophy and (2) computer security research. We develop criteria for computer security-themed trolley problems. We present three such trolley problems (Scenarios~A, B, and~C) and then evaluate those trolley problems under today's main ethical frameworks. 
We provide additional scenarios that, together, surface additional points of consideration about ethics for the computer security research community. 
Given the findings of our research, we reflect and offer considerations for the computer security research community.

\section*{Acknowledgements}

This work was supported in part by the U.S.\ National Science Foundation under awards CNS-2205171 and CNS-2206865, 
the German Federal Ministry of Science and Education (BMBF) as part of the project ``Louisa'' (grant nr 16SV8552), 
the University of Washington Tech Policy Lab (which receives support from the William and Flora Hewlett Foundation, the John D. and Catherine T. MacArthur Foundation, Microsoft, and the Pierre and Pamela Omidyar Fund at the Silicon Valley Community Foundation), and gifts from Google, Meta, Qualcomm, and Woven Planet.
We are grateful to everyone who contributed to this project. Thank you to all who offered comments, questions, insights, and conversations, hosted talks, and reviewed preliminary drafts, including
Lujo Bauer,
Hauke Behrendt,
Dan Boneh, 
Kevin Butler,
Aylin Caliskan,
Inyoung Cheong,
Lorrie Faith Cranor,
Sauvik Das,
Zakir Durumeric,
Kevin Fu,
Alex Gantman,
Gennie Gebhart,
Kurt Hugenberg,
Umar Iqbal,
Apu Kapadia,
Erin Kenneally,
David Kohlbrenner,
Seth Kohno,
Phil Levis,
Rachel McAmis,
Alexandra Michael,
Bryan Parno,
Elissa Redmiles, 
Katharina Reinecke, 
Franziska Roesner,
Stefan Savage,
Stuart Schechter,
Sudheesh Singanamalla,
Patrick Traynor,
Emily Tseng, and
Miranda Wei.
We thank the anonymous USENIX Security 2023 reviewers for their insightful feedback, comments, and suggestions.
We also sincerely thank all attendees of past presentations about this work.

\balance

\appendix

\section{Additional Scenario Contributions}
\label{ap:scenarios-more-full}

Here we provide more details about the scenarios mentioned in Section~\ref{sec:scenarios-more}.

Unlike the trolley problem-like scenarios from Section~\ref{sec:scenarios-initial}, for the scenarios here, we loosen the design criteria and no longer require the scenarios to be ``early'', have ``binary decision'', be ``dilemmas'', and yield ``decisive'' analyses.
Instead, our primary goals are for these new scenarios to (1) reflect scenarios encountered in the computer security research field and (2) inspire thoughtful conversation and reflection upon how individuals and the computer security research field at large reason about moral decisions.

We assess goal (1) by construction and through reflection upon  our own experiences as computer security researchers and, additionally, through the review of our scenarios with others security researchers. Goal (2) is more subjective; we highlight specific points for consideration and reflection in our discussions below.

\subsection{Vulnerability Disclosure Scenarios}
\label{ap:scenario:disclosure}

After discovering a vulnerability in a company's product, computer security researchers often face the following question: do they disclose privately to the company first, before publicly discussing their results (e.g., before publishing a paper)? Or, do they instead publicly discuss their results first, before privately disclosing to the company?

Scenario A (Section~\ref{sec:scenario:medical} and Figure~\ref{fig:scenario:medical_device} in Appendix~\ref{ap:figures}) captures a  complicated vulnerability disclosure situation in which it is impossible to patch the vulnerable artifact because the maker of the artifact no longer exists. In general, there \emph{is} a company to disclose to, it \emph{is} possible to patch the vulnerable artifact, and the company \textit{will} patch the vulnerability once it is disclosed to them.
Scenario~D1 provides an example of this general setting (Figure~\ref{fig:scenario:disclosure:base} in Appendix~\ref{ap:figures}).

By itself, Scenario~D1 is rather uninteresting: coordinated disclosure best practices in the computer security research community would have the researchers  privately disclose the vulnerability to the company before any public discussion. We present Scenario~D1 not because it is interesting unto itself but because, as a baseline, it enables the philosophical exploration of variants to Scenario~D1 where the ethical analysis might be different or more complicated.

We summarize Scenario~D1 and the sequence of related scenarios below, beginning first with more details about Scenario~D1:

\begin{itemize}
\item \textbf{Scenario D1, Vulnerability Disclosure (Base Case).}  Scenario~D1 (Figure~\ref{fig:scenario:disclosure:base} in Appendix~\ref{ap:figures}) is a generic vulnerability disclosure scenario in which researchers discover a vulnerability in a product made by Company~D. 

The description of Scenario~D1 contains additional complexities; these complexities exist in order to enable precise comparisons between Scenario~D1 and  the other scenarios below.

Company~D is known \textit{not} to be litigious. 
Company~D is known to take security seriously and will begin working on a patch as soon as it learns about a vulnerability. The researchers are tenured full professors who do not need a publication. The timeline for adversaries to manifest without a public vulnerability disclosure is one year; the timeline for how long it will take to develop a patch once Company D knows about the vulnerability is six months; and it will take adversaries three months to weaponize the vulnerability if it were publicly disclosed first. Given these timelines, a private vulnerability disclosure to Company D would give it enough time to patch its product before adversaries manifest. On the other hand, a public disclosure first would make users vulnerable to significant harm for three months (adversaries manifest after three months, a patch is available after six months). 

\end{itemize}

\begin{itemize}
\item \textbf{Scenario D2, Vulnerability Disclosure (Legal Threat).} Scenario~D2 (Figure~\ref{fig:scenario:disclosure:legalthreat} in Appendix~\ref{ap:figures}) is a variant of Scenario~D1 in which Company~D is known to be highly litigious and to not consider security seriously. If the researchers disclose privately to the company first, the researchers would become entangled in a legal battle and not be able to publicly disclose the vulnerability. Further, the company will not begin efforts to patch their product. This means that when adversaries manifest (and they are guaranteed to manifest in a year), users will be significantly harmed. It is only after the public learns about the vulnerability, e.g., via a public disclosure or the manifestation of adversaries, that the company will begin to develop a patch.

A public disclosure would thus result in users being vulnerable for three months (starting after month three and until after month six). A private disclosure to Company D would result in users being vulnerable for six months (starting after one year and continuing until after eighteen months). 

\item \textbf{Scenario D3, Vulnerability Disclosure (Legal Threat with Uncertainty).} Scenario~D3 (Figure~\ref{fig:scenario:disclosure:legalthreat:uncertainty} in Appendix~\ref{ap:figures}) is like Scenario~D2 except now there is uncertainty whether Company D is highly litigious and whether they take security seriously. It \emph{is} known that Company D is in an industry (Industry~D) that is highly litigious and that does not take security seriously. 

Compared to Scenario~D2, the central ethical question is thus whether the researchers should assume (with high probability) that Company D is like the rest of Industry~D or whether to give Company~D the benefit of the doubt and assume that it is like any other unknown company.

\item \textbf{Scenario D4, Vulnerability Disclosure (Legal Threat with Uncertainty and Uncertain Adversaries).} Scenario~D4 (Figure~\ref{fig:scenario:disclosure:legalthreat:uncertainty:uncertainadvesaries} in Appendix~\ref{ap:figures}) is like Scenario~D3 but adds even greater uncertainty. Whereas the researchers in Scenario~D3 know for a fact that adversaries would manifest in a year, the researchers in this scenario do not know when (or even if) adversaries will manifest if there is no public disclosure.

\item \textbf{Scenario D5, Vulnerability Disclosure (Legal Threat, Security Responsible).} 
Scenario~D5 (Figure~\ref{fig:scenario:disclosure:legalthreat:securityresponsible} in Appendix~\ref{ap:figures}) is like Scenario~D2 except that it separates the legal battle following a private vulnerability disclosure  from the protection of users.
In this scenario, the day before their planned publication date, the researchers learn that the company \textit{will} start taking computer security vulnerabilities seriously. Given this change in policy for Company~D, if the researchers privately disclose to the company first, they will not be able to publish their results for three years (Company~D remains litigious), but the company will begin working on a patch and users will be protected from any eventual manifestation of adversaries. The researchers are all tenured full professors who would not be significantly harmed if they are unable to publish for three years.

\item \textbf{Scenario D6, Vulnerability Disclosure (Legal Threat, Security Responsible, Career-critical Research).} Scenario~D6 (Figure~\ref{fig:scenario:disclosure:legalthreat:securityresponsible:student} in Appendix~\ref{ap:figures}) is like Scenario~D5 except that the lead researcher is  a senior PhD student and the planned publication is their final PhD defense and the filing of their dissertation. They will be unable to defend and graduate if their research (the final portion of their dissertation) becomes entangled in a legal battle. If they cannot defend and file their dissertation, then they must decide whether to decline a job offer and remain in graduate school for longer or leave academia without a PhD. 
The PhD student's department's executive committee met and decided that \dash unless the student's PhD advisor intervenes \dash the student should defend and file their dissertation as planned and \emph{not} notify Company~D first. The department chair additionally told the PhD advisor that it is their responsibility \dash not the student's \dash to decide whether to disrupt the current plans and notify Company~D before the defense and dissertation filing.

\item \textbf{Scenario D7, Vulnerability Disclosure (Legal Company, Law-breaking Users).} Scenario~D7 (Figure~\ref{fig:scenario:disclosure:legalthreat:usersbreaklaw} in Appendix~\ref{ap:figures}) is like Scenario~D1 in Figure~\ref{fig:scenario:disclosure:base} in that Company~D is not litigious and takes the development of security patches seriously. It is unlike the  company in Scenario~D1 in that this scenario's Company~D, although operating legally in a country due to a legal loophole, is used by people who are breaking the law. Specifically, the users with accounts on Company~D's web service use Company~D's web server to share non-consensual explicit material (often called revenge porn), an act that is illegal.

\end{itemize}

\subsection{Paper Raises Concerns}
\label{ap:scenario:submissionconcerns}

We now turn to another set of scenarios, beginning with the base scenario, Scenario~E1 (Figure~\ref{fig:scenario:program_committee_concerns:base}  in Appendix~\ref{ap:figures}). In Scenario~E1, while reviewing a paper under submission, the program committee determines that part of the research (the research reported in Section 9.3 of the submission) should not have been done. Section 9.3 of the submission presents the results of repeated Internet-wide scans. The program committee observes that the scanning infrastructure could have caused computers, including insulin pumps in hospitals, to crash. Moreover, after discussing with staff at a local hospital, the program committee learns that the researcher's Internet-wide scans \emph{did} cause insulin pumps to crash. 

We describe key elements of Scenario~E1, as well as the subsequent scenarios in this collection of scenarios, below:

\begin{itemize}

\item \textbf{Scenario~E1 (Base Case).} In this base case scenario (Figure~\ref{fig:scenario:program_committee_concerns:base} in Appendix~\ref{ap:figures}), program committee members are only given two options: (1) to reject the paper or (2) to accept the paper as-is, with no required modifications. If option (2) is selected, the authors will receive reviews from the program committee and will have the opportunity to revise the paper however they see fit.

\item \textbf{Scenario~E2 (Reject, Accept, Remove).} This scenario builds on Scenario~E1. In this scenario (Figure~\ref{fig:scenario:program_committee_concerns:moreoptionsremove} in Appendix~\ref{ap:figures}),  program committee members are given an additional option: (3) to accept the paper under the condition that the authors agree to remove Section 9.3.

\item \textbf{Scenario~E3 (Accept, Reject, Remove, Note).} This scenario builds on Scenario~E2. In this scenario (Figure~\ref{fig:scenario:program_committee_concerns:moreoptions} in Appendix~\ref{ap:figures}),  program committee members are given an additional option: (4) to accept the paper under the condition that the authors agree to add a clearly visible note to the first page, written by the program committee, detailing the ethical concerns with the methods used in Section 9.3.

\item \textbf{Scenario~E4 (Extensive Author Testing).} This scenario builds on Scenario~E3. In this scenario (Figure~\ref{fig:scenario:program_committee_concerns:authorstried}  in Appendix~\ref{ap:figures}), the program committee learns that the authors conducted extensive, thorough testing and evaluation prior to deploying their crawling infrastructure and believed that their scans would not cause any crashes. Out of an abundance of caution, the authors also developed mechanisms to learn whether their system caused any crashes. For example, the authors followed the practices described in~\cite{durumeric2013zmap}. The reason the authors did not learn about crashes to insulin pumps was because of errors made by the operators of the impacted insulin pumps.

\item \textbf{Scenario~E5 (Known Risk and Moral Responsibility).} This scenario builds on Scenario~E3. In this scenario (Figure~\ref{fig:scenario:program_committee_concerns:authorsknewandfeltduty} in Appendix~\ref{ap:figures}), after extensive testing, the researchers learned that their crawling infrastructure could cause some computers to crash. They decided to proceed with their Internet-wide scans anyway, out of a sense of duty and moral responsibility. Because of the seriousness of the vulnerability, they felt a need to crawl the Internet, identify vulnerable webservers, and then contact those webservers' operators and provide instructions on how to patch. In fact, there is a direct and measurable impact to the security of those webservers (and their users) because of the authors' crawls and subsequent efforts to reach webserver operators. Due to space limitations, the authors did not include any of the above details in their submission \dash the program committee members only learned these details because the program chairs reached out to the authors with questions.

\item \textbf{Scenario~E6 (Authors Ignored Risks)}. This scenario builds on Scenario~E3. In this scenario (Figure~\ref{fig:scenario:program_committee_concerns:authorsknew} in Appendix~\ref{ap:figures}), the authors knew that their scans could cause crashes but proceeded anyway because they thought that the results of their scans would increase the likelihood of their paper being accepted. The authors believed that a few crashes here and there would be okay since people's computers crash all the time anyway.

\item \textbf{Scenario~E7 (Authors Ignored Risks, Luck Prevents Crashes)}. This scenario builds on Scenario~E6.
In this scenario (Figure~\ref{fig:scenario:program_committee_concerns:authorsknew:nocrashesluck} in Appendix~\ref{ap:figures}), the crawling infrastructure did \emph{not} cause any crashes out of sheer luck \dash the researchers accidentally had a bug in their crawling infrastructure that caused extra delays, and the extra delays meant no crashes. Nevertheless, as with Scenario~E6, the researchers believed that their crawls could cause crashes and proceeded anyway because they thought that the results of their scans would increase the likelihood of their paper being accepted. The extra delays in the crawling infrastructure do not impact the correctness of the results in Section 9.3.

\item \textbf{Scenario~E8 (Authors Ignored Risks, Section 9.3 Results Critical).} This scenario builds on Scenario~E6, where the researchers knew that their scans could cause crashes but proceeded anyway because they thought that the results of their scans would increase the likelihood of their paper being accepted.
In this scenario (Figure~\ref{fig:scenario:program_committee_concerns:authorsknew:73critical} in Appendix~\ref{ap:figures}), the program committee believes that it is vital for the results in Section 9.3 to be published. With the publication of the paper, and the results in Section 9.3, the remaining vulnerable webservers will have increased motivation to patch. Without the publication of the paper, including Section 9.3, many webservers and hence many users will remain vulnerable.

\item \textbf{Scenario~E9 (Authors Ignored Risks, Moral Implications Not Realized).} This scenario builds on Scenario~E6.
In this scenario (Figure~\ref{fig:scenario:program_committee_concerns:authorsknew:phdstudent:implicationsnotrealized} in Appendix~\ref{ap:figures}), the researchers are deeply regretful after they are informed by the program chair about the crashes caused by their crawls. With this new information, the researchers strongly wish that they did not conduct the crawls reported in Section 9.3.

\end{itemize}

For these scenarios, the program committee will be evaluating not just the morality of the work done, but their moral responsibilities as a program committee.

\subsection{After Rejection}
\label{ap:scenario:afterrejection}

Our Scenario~F continues Scenario~E1 from the perspective of the researchers: after review by the program committee, the program committee rejected the authors' paper.  The rejection email explains to the authors the reason for the rejection: that the authors' crawls caused medical devices to crash. Until receiving the notification email, the authors did not know that their crawls could crash machines. But, now that they do know, they agree that the research in Section 9.3 should not have been done. The authors believe that their research, even without Section 9.3, is valuable and important to publish. The authors know of another conference with zero overlapping program committee members with the conference that rejected the authors' research, and the authors know that the two conferences' program committees will not discuss. (Scenario~F is captured in Figure~\ref{fig:scenario:resubmit_paper_with_concerns} in Appendix~\ref{ap:figures}.)

The question for these researchers is: what should they do? Should they stop working on the project and never try to publish any part of it? Should they submit the paper again, to the new conference, without modification? Should they remove Section 9.3 and, in their new submission, pretend like Section 9.3 never existed and that the crawls never happened? Or should they keep Section 9.3 in their new submission, but add a note about what happened and why, in retrospect, they should not have done those crawls?

\section{Figures for Scenarios}
\label{ap:figures}

Since the main body of the text captures key elements of our scenarios, for readability, we do not include our scenario figures in the body of the paper. The scenario figures are in this appendix. The following is a mapping from scenarios to figures:

\begin{itemize}
\item \textbf{Scenario A}: Figure~\ref{fig:scenario:medical_device}. (Section~\ref{sec:scenario:medical}.)

\item \textbf{Scenario B}: Figure~\ref{fig:scenario:immoral_data_jobs}. (Section~\ref{sec:scenario:immoraldata}.)

\item \textbf{Scenario C}: Figure~\ref{fig:scenario:inadvertent_disclosure}. (Section~\ref{sec:scenario:inadvertentdisclosure}.)

\item \textbf{Scenario D1}: Figure~\ref{fig:scenario:disclosure:base}. 
(Section~\ref{sec:scenarios:more:disclosure}.)

\item \textbf{Scenario D2}: Figure~\ref{fig:scenario:disclosure:legalthreat}. 
(Section~\ref{sec:scenarios:more:disclosure}.)

\item \textbf{Scenario D3}: Figure~\ref{fig:scenario:disclosure:legalthreat:uncertainty}. 
(Section~\ref{sec:scenarios:more:disclosure}.)

\item \textbf{Scenario D4}: Figure~\ref{fig:scenario:disclosure:legalthreat:uncertainty:uncertainadvesaries}. 
(Section~\ref{sec:scenarios:more:disclosure}.)

\item \textbf{Scenario D5}: Figure~\ref{fig:scenario:disclosure:legalthreat:securityresponsible}. 
(Section~\ref{sec:scenarios:more:disclosure}.)

\item \textbf{Scenario D6}: Figure~\ref{fig:scenario:disclosure:legalthreat:securityresponsible:student}. 
(Section~\ref{sec:scenarios:more:disclosure}.)

\item \textbf{Scenario D7}: Figure~\ref{fig:scenario:disclosure:legalthreat:usersbreaklaw}. 
(Section~\ref{sec:scenarios:more:disclosure}.)

\item \textbf{Scenario E1}: Figure~\ref{fig:scenario:program_committee_concerns:base}.
(Section~\ref{sec:scenarios:more:submission}.)

\item \textbf{Scenario E2}: Figure~\ref{fig:scenario:program_committee_concerns:moreoptionsremove}. 
(Section~\ref{sec:scenarios:more:submission}.)

\item \textbf{Scenario E3}: Figure~\ref{fig:scenario:program_committee_concerns:moreoptions}. 
(Section~\ref{sec:scenarios:more:submission}.)

\item \textbf{Scenario E4}: Figure~\ref{fig:scenario:program_committee_concerns:authorstried}. 
(Section~\ref{sec:scenarios:more:submission}.)

\item \textbf{Scenario E5}: Figure~\ref{fig:scenario:program_committee_concerns:authorsknewandfeltduty}. 
(Section~\ref{sec:scenarios:more:submission}.)

\item \textbf{Scenario E6}: Figure~\ref{fig:scenario:program_committee_concerns:authorsknew}. 
(Section~\ref{sec:scenarios:more:submission}.)

\item \textbf{Scenario E7}: Figure~\ref{fig:scenario:program_committee_concerns:authorsknew:nocrashesluck}. 
(Section~\ref{sec:scenarios:more:submission}.)

\item \textbf{Scenario E8}: Figure~\ref{fig:scenario:program_committee_concerns:authorsknew:73critical}. 
(Section~\ref{sec:scenarios:more:submission}.)

\item \textbf{Scenario E9}: Figure~\ref{fig:scenario:program_committee_concerns:authorsknew:phdstudent:implicationsnotrealized}. 
(Section~\ref{sec:scenarios:more:submission}.)

\item \textbf{Scenario F}: Figure~\ref{fig:scenario:resubmit_paper_with_concerns}. 
(Section~\ref{sec:scenarios:more:rejection}.)

\end{itemize}

For Figure~\ref{fig:scenario:medical_device}, the 85\,000 number of patients  corresponds to roughly 2.6 in every 10\,000 people, which is one-tenth the rate of the number of people who had pacemakers as of 1988~\cite{NumberIMDs}. For the calculation of 85\,000, we used the U.S.\ Census Bureau's estimated population on January 1, 2022~\cite{Census2022}.

\begin{figure*}
\begin{tcolorbox}
\begin{center}

\textbf{Computer Security Scenario A: Medical Device Vulnerability}\end{center}

\textbf{Context:}
\begin{itemize}[itemsep=0pt, parsep=2pt]
\item Company A produces a lifesaving wireless implantable medical device. It is the only device of its type ever invented. When a patient receives this device, it will (on average) extend their lifespan by ten years.
\item 
Company A goes bankrupt and closes due to poor financial practices, including a failure to calculate the market size and the costly manufacture of hundreds of thousands of devices before they were  needed.
\item 
At the time of Company A's bankruptcy, approximately 85\,000 people in the United States use Company A's device, and many more people globally.
\item Doctors continue to implant the surplus of (now unsupported) device in new patients.
\item 
Shortly after Company~A closes, researchers discover a software vulnerability in the device. If exploited, the vulnerability could cause significant harm to the patients. Since Company A no longer exists, the software cannot be updated to address this vulnerability.
\item
The researchers know that there is \textit{zero} probability that the vulnerability will ever be exploited \emph{even if} the vulnerability is disclosed to the public.
\item The computer security research field and the healthcare industry have already internalized the importance of computer security for wireless implantable medical devices; there are no field- or industry-wide gains to be made by disclosing the vulnerabilities to the public.
\end{itemize}

\textbf{The choice for the researchers:}
\begin{itemize}[itemsep=0pt, parsep=2pt]
    \item \textit{Not disclose the vulnerability to anyone:} Patients will have no awareness that their device is vulnerable; patients will keep and / or proceed with obtaining the device and receive significant health benefits.
    \item \textit{Disclose the vulnerability to the healthcare industry, patients, and the public:} Patients will have the choice to remove or not receive the device; there is a risk of health harm to patients if patients remove and / or do not receive the device; there is a risk of psychological harm to patients and loved ones if patients know that they have a vulnerable device in their bodies (even if they also are told that the likelihood of compromise is  zero); given the psychological harms, most patients would have preferred not to have learned about the vulnerability.
\end{itemize}
\end{tcolorbox}
\caption{A computer security scenario in which vulnerabilities are found in an unsupported medical device. See Section~\ref{sec:scenario:medical}.}
\label{fig:scenario:medical_device}

\end{figure*}

\begin{figure*}
\begin{tcolorbox}
\begin{center}

\textbf{Computer Security Scenario B: Research with Stolen Data}\end{center}

\textbf{Context:}
\begin{itemize}[itemsep=0pt, parsep=2pt]
\item Company B offers a service to connect employers (jobs) with applicants. Employers submit a job description. Applicants submit their resumes and answer job-specific questions. Company~B's AI system selects the ``best'' candidates from the set of applicants and forwards only those selected applicants to employers.
\item Applicants may consider some or all of their application packets as confidential, e.g., some applicants may not wish for their current employers to know that they are searching for a new job or  searched for a new job in the past, and some job-specific questions may have answers that applicants do not wish to be public (e.g., questions about individual strengths and weaknesses, or questions about why they are applying for a new job).
\item Analysts predict that in five years, Company~B will process one hundred times more applicants per month than it does today. 
\item There is public suspicion but not proof that Company~B's AI system exhibits significant biases, e.g., biases based on race and gender, and that these biases result in job applications from members of marginalized populations \emph{not} being forwarded to employers as frequently as applications from members of non-marginalized groups. There is also public suspicion but not proof that Company~B's AI systems may be vulnerable to adversarial manipulation.
\item Company B was compromised and the entirety of their data was stolen. This data includes data on all jobs ever posted, all application packets ever submitted, all past results of the job-applicant matching AI system, and all information about the internal workings of the company's AI system (ML models, etc.).

\item Researchers wish to study the stolen data. 
The research will provide a concrete, scientific assessment of whether Company~B's AI system is biased.
Given the entirety of Company~B's data, the researchers can assess the past impacts of such biases, e.g., the researchers can count the number of application packets \emph{not} forwarded to employers as a result of AI bias. Using their findings, the researchers can propose methods to reduce biases.
\item If biases are present in the AI system, then removing those biases would result in a change in which applicants are shown to employers (assuming the number of applicants shown to an employer is constant). For this scenario, assume that anyone removed from that set of applicants will still be able to find a job that they desire.
\item The researchers know that adversaries will also study the stolen data. Hence, the researchers will study the potential for adversaries to exploit the AI system, including its internal biases, to undeservedly match with a job. The researchers will propose mitigations to any vulnerabilities that they find.
\item The thieves of Company~B's data posted all the stolen materials online.
\item Many job applicants publicly stated their desire for the stolen data to be permanently deleted, everywhere. 
The researchers obtained a copy of the stolen data as soon as it became available online, before the job applicants stated their desires for the data to be deleted and before all publicly-available copies were deleted.
If the researchers do their research, they will need to retain a copy of the data in case their results are challenged (e.g., the company challenges the results).
\item If the researchers study the data and report on their results, they know not to include anything in their publication that could lead to the identification of any of the applicants (e.g., no direct quotes from resumes, no job titles and company names of people who applied for jobs).
\end{itemize}

\textbf{The choice for the researchers:}
\begin{itemize}[itemsep=0pt, parsep=2pt]
\item 
\textit{Not study the data:} People whose data was stolen will appreciate that the researchers did not further violate their rights to privacy and informed consent.
\item
\textit{Study the data:} The results will uncover whether Company~B's AI systems are biased and / or vulnerable to manipulation and, if so, provide a foundation for mitigating such risks with future versions of Company~B's and other job-applicant matching systems, including possibly minimizing the use of AIs.
\end{itemize}
\end{tcolorbox}
\caption{A computer security scenario in which researchers wish to study stolen data.
See Section~\ref{sec:scenario:immoraldata}.}
\label{fig:scenario:immoral_data_jobs}
\end{figure*}

\begin{figure*}
\begin{tcolorbox}
\begin{center}

\textbf{Computer Security Scenario C: Inadvertent Vulnerability ``Disclosure''}\end{center}

\textbf{Context:}
\begin{itemize}[itemsep=0pt, parsep=2pt]
\item
A research paper is submitted to a top peer-reviewed conference; the paper details the discovery of a previously unknown vulnerability in the product from Company C.
\item
The vulnerability has not yet been disclosed to Company~C.
\item 
The authors in their submission write that do not want to disclose to Company C until after the paper has been officially accepted. The authors' reasons are not articulated in the paper.
\item 
A person from Company C is on the program committee.
\item 
That person reads the paper and realizes that the vulnerability can lead to serious harms to millions of users if exploited.
\item
That person believes that it will take significant time and effort to patch the vulnerability.
\item 
The program committee chairs required all program committee members to explicitly agree to maintain the confidentiality of submissions and not discuss submissions with anyone outside of the program committee.
\item Before agreeing to join the program committee, the employee of Company C sought permission from Company C's leadership team; Company C's leadership team understood the responsibilities of  committee members and agreed to let the employee join.
\end{itemize}

\textbf{The choice for the employee of Company C:}
\begin{itemize}[itemsep=0pt, parsep=2pt]
\item 
\textit{Not report the vulnerability to Company C:} Confidentiality of program committee process preserved; authors have agency over when and how they disclose the vulnerability to Company~C.
\item 
\textit{Report the vulnerability to Company C}: Company C can check to see if the vulnerability is already being exploited; Company C can immediately begin working on solutions to protect their users; users’ security, privacy, and safety will be protected as early as possible.
\end{itemize}
\end{tcolorbox}
\caption{A computer security scenario in which a program committee member reads a confidential paper that presents an undisclosed vulnerability in the software produced by the program committee member's company.
See Section~\ref{sec:scenario:inadvertentdisclosure}.}
\label{fig:scenario:inadvertent_disclosure}

\end{figure*}

\begin{figure*}
\begin{tcolorbox}
\begin{center}
\textbf{Computer Security Scenario D1: Vulnerability Disclosure (Base Case)}\end{center}
\textbf{Context:}
\begin{itemize}[itemsep=0pt, parsep=2pt]
\item Researchers discover a vulnerability in Company D’s product; the researchers must decide whether or not to disclose the vulnerability to Company D before their research paper is published.
\item Once Company D makes the decision to fix the vulnerability, it will take them six months to complete the process and release a patch.
\item Once adversaries learn about the vulnerability, it will take them three months to weaponize the vulnerability, after which the weapon will be deployed.
\item Once the cyber weapon is deployed, each of Company D’s users are at risk of losing 25\% of their retirement savings; there is no way for users to move their retirement savings into other systems (they are locked into using Company D’s product); Company D has ten million users;  15\% of users will be impacted during each month of vulnerability (until the system is patched).
\item The researchers got the inspiration for their vulnerability research from monitoring chatter on underground forums; given that chatter, the researchers believe that adversaries will independently discover the vulnerability in nine months and deploy a cyber weapon in one year.
\item The researchers believe that Company D will be responsible: Company D will immediately begin working on a patch after a private vulnerability disclosure and will not entangle the researchers in a legal battle aimed at preventing the publication of the research paper. Further, the researchers are confident that the entirety of their research process was legal.
\item The researchers are all tenured full professors who, from a career perspective, do not need a publication.
\item The publication program committee has no stated preference on what the authors do; they trust  authors to make the right decision.
\end{itemize}
\textbf{The choice for the researchers:}
\begin{itemize}[itemsep=0pt, parsep=2pt]
\item \textit{Disclose the vulnerability to Company D and wait six months before publishing their paper:} Since Company D is believed to be responsible, the company will immediately begin working on a patch. Six months later, after the patch is deployed, the researchers can publish their paper and Company D’s users will be secure against the vulnerability.
\item \textit{Do not disclose the vulnerability to Company D before publishing their paper:} Once the paper is published (month zero), adversaries will start to weaponize the vulnerability and the company will start working on a patch. Adversaries will deploy their cyber weapon after three months (month three); the patch will be deployed after six months (month six); this situation leaves \textit{three months} in which Company D’s users are actively being exploited (months four, five, and six).
\end{itemize}
\end{tcolorbox}
\caption{A computer security scenario in which researchers have discovered a vulnerability in a product that can be patched and in which the vulnerable company is believed to be responsible.}
\label{fig:scenario:disclosure:base}
\end{figure*}

\begin{figure*}
\begin{tcolorbox}
\begin{center}
\textbf{Computer Security Scenario D2: Vulnerability Disclosure (Legal Threat)}\end{center}
\textbf{Context:} Equivalent to Scenario D1 in Figure~\ref{fig:scenario:disclosure:base} except:
\begin{itemize}[itemsep=0pt, parsep=2pt]
\item Company D is known to be highly litigious. If the researchers  disclose the vulnerability first to Company D, the researchers will be drawn into a legal battle and be unable to publicly discuss their findings and the vulnerability for at least three years.
This legal battle will happen even though the researchers are confident that the entirety of their research process was legal.
\item Company D is known to not take computer security seriously unless there is an incident or public pressure. A private vulnerability disclosure to Company D will not cause it to begin working on defenses; it will only begin working on defenses after the vulnerability is actively exploited or there is public pressure.
\item The researchers do not fear legal action against themselves if they were to publish their paper before sharing the vulnerability with the company.
\end{itemize}
\textbf{The choice for the researchers:}
Equivalent to Scenario D1 in Figure~\ref{fig:scenario:disclosure:base} except:
\begin{itemize}[itemsep=0pt, parsep=2pt]
\item \textit{Disclose the vulnerability to Company D before publishing their paper and become entangled in a legal battle:} The researchers will not be able to publish their findings for at least three years. Adversaries will manifest in one year, after which Company D will begin working on a patch, which would not be released for another six months. 
This situation leaves \textit{six months} in which Company D's users are actively being exploited (months thirteen to eighteen).
\end{itemize}
\end{tcolorbox}
\caption{A computer security scenario like Scenario D1 in Figure~\ref{fig:scenario:disclosure:base} except that the company producing the product is not responsible: the company will entangle the researchers in a legal battle, prevent publication, and not begin working on a security patch after receiving the private vulnerability disclosure.}
\label{fig:scenario:disclosure:legalthreat}
\end{figure*}

\begin{figure*}
\begin{tcolorbox}
\begin{center}
\textbf{Computer Security Scenario D3: Vulnerability Disclosure (Legal Threat with Uncertainty)}\end{center}
\textbf{Context:} Equivalent to Scenario D2 in Figure~\ref{fig:scenario:disclosure:legalthreat} except:
\begin{itemize}[itemsep=0pt, parsep=2pt]
\item The security field does not have any direct experience with vulnerability disclosures to Company D.
\item Company D is in Industry D.
\item  Other companies in Industry D are highly litigious. If the researchers disclose the vulnerability first to Company D, and if Company D is like the other companies in Industry D, the researchers will be drawn into a legal battle and be unable to publicly discuss their findings and the vulnerability for at least three years. This legal battle would happen even though the researchers are confident that the entirety of their research process was legal.
\item Other companies in Industry D do not take computer security seriously unless there is an incident or public pressure. If Company D is like the other companies in Industry D, a private vulnerability disclosure to Company D will not cause it to begin working on defenses;  it will only begin working on defenses after the vulnerability is actively exploited or there is public pressure.
\end{itemize}
\textbf{The choice for the researchers:}
Equivalent to Scenarios D1 
in Figure~\ref{fig:scenario:disclosure:base}
and Scenario D2 in Figure~\ref{fig:scenario:disclosure:legalthreat} except:
\begin{itemize}[itemsep=0pt, parsep=2pt]
\item \textit{Disclose the vulnerability to Company D before publishing their paper:}
If Company D is like other companies in Industry D, the situation is the same as in Scenario D2 and Figure~\ref{fig:scenario:disclosure:legalthreat}.
If Company D is like companies in \textit{other industries}, the situation is the same as in Scenario D1 and Figure~\ref{fig:scenario:disclosure:base}.
\end{itemize}
\end{tcolorbox}
\caption{A computer security scenario like Scenario D2 in Figure~\ref{fig:scenario:disclosure:legalthreat} except that the company producing the product is in an industry known to be irresponsible: other companies in the industry would  entangle the researchers in a legal battle, prevent publication, and not begin working on a security patch after receiving the vulnerability disclosure. However, it is not known whether the company itself is responsible or not.}
\label{fig:scenario:disclosure:legalthreat:uncertainty}
\end{figure*}

\begin{figure*}
\begin{tcolorbox}
\begin{center}
\textbf{Computer Security Scenario D4: Vulnerability Disclosure (Legal Threat with Uncertainty and Uncertain Adversaries)}\end{center}
\textbf{Context:} Equivalent to Scenario D3 in Figure~\ref{fig:scenario:disclosure:legalthreat:uncertainty} except:
\begin{itemize}[itemsep=0pt, parsep=2pt]
\item The researchers know that adversaries are always finding surprising, novel, new vulnerabilities in systems.
\item 
However, the researchers are unable to predict when, if ever, adversaries will independently discover and weaponize the vulnerability in question with respect to Company D’s product. The researchers do, however, know a lower-bound on when adversaries would manifest without a public disclosure: without a public disclosure, it will be \emph{at least} nine months before adversaries discover the vulnerability and at least one year before the adversaries have a weaponized exploit.
\end{itemize}
\textbf{The choice for the researchers:}
Equivalent to Scenario D3 
in Figure~\ref{fig:scenario:disclosure:legalthreat:uncertainty} except:
\begin{itemize}[itemsep=0pt, parsep=2pt]
\item \textit{Disclose the vulnerability to Company D before publishing their paper:}
If Company D is like other companies in Industry D, the situation is the same as in Scenario D2 and Figure~\ref{fig:scenario:disclosure:legalthreat} \textit{except} that it is unknown if or when adversaries will manifest and hence it is unknown whether Company D's users will be actively compromised or not after one year.
If Company D is like companies in other industries, the situation is the same as in Scenario D1 and Figure~\ref{fig:scenario:disclosure:base}.
\end{itemize}
\end{tcolorbox}
\caption{A computer security scenario like Scenario D3 in Figure~\ref{fig:scenario:disclosure:legalthreat:uncertainty} except that the researchers cannot predict when adversaries will independently find and start weaponizing the vulnerability.}
\label{fig:scenario:disclosure:legalthreat:uncertainty:uncertainadvesaries}
\end{figure*}

\begin{figure*}
\begin{tcolorbox}
\begin{center}
\textbf{Computer Security Scenario D5: Vulnerability Disclosure (Legal Threat, Security Responsible)}\end{center}
\textbf{Context:} Equivalent to Scenario D2 in Figure~\ref{fig:scenario:disclosure:legalthreat} except:
\begin{itemize}[itemsep=0pt, parsep=2pt]
\item The researchers did an ethics analysis and determined that, given Company~D's litigious nature and their lack of concern for computer security, the morally right decision is to \textit{not} discuss the vulnerability with Company D before publishing their research results.
\item The day before the researchers plan to publish their results, they learn that Company~D's internal policy has changed and that, starting now, they \emph{do} take security seriously; they will begin to develop a patch immediately after receiving notification of a vulnerability, even if information about the vulnerability is not known to the public and even if adversaries are not actively exploiting the vulnerability. The company remains highly litigious.
\end{itemize}
\textbf{The choice for the researchers:}
Equivalent to Scenario D2 in Figure~\ref{fig:scenario:disclosure:legalthreat} except:
\begin{itemize}[itemsep=0pt, parsep=2pt]
\item \textit{Disclose the vulnerability to Company D before publishing their paper; become entangled in a legal battle and be unable to publish:} The researchers will not be able to publish their findings for at least three years. Recall from the description of Scenario~D1 in Figure~\ref{fig:scenario:disclosure:base} that the researchers are all tenured full professors who, from a career perspective, do not need a publication. Company D will begin working on a patch immediately and users will be protected after the patch is deployed six months later and before any adversaries manifest.
\end{itemize}
\end{tcolorbox}
\caption{A computer security scenario like Scenario D2 in Figure~\ref{fig:scenario:disclosure:legalthreat} except that the researchers know for certain that the company takes security seriously and will immediately begin working on defenses after receiving a vulnerability disclosure.}
\label{fig:scenario:disclosure:legalthreat:securityresponsible}
\end{figure*}

\begin{figure*}
\begin{tcolorbox}
\begin{center}
\textbf{Computer Security Scenario D6: Vulnerability Disclosure (Legal Threat, Security Responsible, Career-critical Research)}\end{center}
\textbf{Context:} Equivalent to Scenario D5 in Figure~\ref{fig:scenario:disclosure:legalthreat:securityresponsible} except:
\begin{itemize}[itemsep=0pt, parsep=2pt]
\item The lead researcher is a senior PhD student. The other author is the student's PhD advisor.
\item The planned publication was the PhD student's final PhD defense and the filing of their dissertation.
\item If the research becomes entangled in a legal battle and the researchers cannot publish their findings, then the lead researcher will not be able to give their final defense and file their dissertation as planned.
\item The lead researcher has been offered an industry job, which they plan to accept; if the researcher cannot defend and cannot file their dissertation, they will need to decide whether to (1) decline the industry job and remain in academia until they can complete another research project (1--2 years) or (2) accept the industry job and leave academia without a PhD.
\item The PhD student's department's executive committee  met and decided that the student should defend and file their dissertation as planned, and \textit{not} notify Company~D before the defense and dissertation filing, unless the advisor intervenes. The department chair additionally tells the advisor that it is the advisor's responsibility \dash not the student's \dash to decide whether to notify Company~D prior to the defense and dissertation filing. If the PhD advisor wishes to intervene and notify Company~D first, they have until the end of the day to do so.
\end{itemize}
\textbf{The choice for the student's PhD advisor:}
Equivalent to Scenario D5 in Figure~\ref{fig:scenario:disclosure:legalthreat:securityresponsible} except:
\begin{itemize}[itemsep=0pt, parsep=2pt]
\item \textit{Disclose the vulnerability to Company D before the PhD defense and dissertation filing; become entangled in a legal battle such that the PhD student can't defend and can't file their dissertation:} Unlike Scenario D5 in Figure~\ref{fig:scenario:disclosure:legalthreat:securityresponsible}, the inability to publicly discuss the work (the final defense and the filing of the dissertation) will negatively impact the lead researcher's career. As with Scenario~D5, Company~D will begin working on a patch immediately and users will be protected after the patch is deployed six months later and before any adversaries manifest.
\end{itemize}
\end{tcolorbox}
\caption{A computer security scenario like Scenario D5 in Figure~\ref{fig:scenario:disclosure:legalthreat:securityresponsible} except that the publication of the research is critical for the researcher's career.}
\label{fig:scenario:disclosure:legalthreat:securityresponsible:student}
\end{figure*}

\begin{figure*}
\begin{tcolorbox}
\begin{center}
\textbf{Computer Security Scenario D7: Vulnerability Disclosure (Legal Company, Law-breaking Users)}\end{center}
\textbf{Context:} Equivalent to Scenario D1 in Figure~\ref{fig:scenario:disclosure:base} except:
\begin{itemize}[itemsep=0pt, parsep=2pt]
\item Company D offers a web service that allows  users with accounts to share content, including photos.
\item  
Company D's reason for existence is to facilitate the sharing of non-consensual explicit material (often called revenge porn);
all employees of Company D know that Company D's service is used to share non-consensual explicit material; anyone who visits Company D's web service knows that the web service was designed to facilitate the sharing of non-consensual explicit material.
\item Because Company D's terms of service states that it provides a general web service that simply does not examine or filter content, it is operating legally (even if immorally) within a country.
\item
The only reason a person would create an account for themselves with Company D is if they intend to share non-consensual explicit material; accessing Company D's content does not require an account.
\item However unlikely, it is possible for an attacker to create an account with Company D using someone else's name and email address (e.g., if the attacker first compromises that someone else's email account).
\item When a user with an account uploads non-consensual explicit material to Company D's web service, they are breaking the law. 
\item Once the cyber weapon against Company D’s product is available and if the vulnerability has not been patched, adversaries against Company D can use the weapon to discover the content sharing histories and true identities, including names and email addresses, of all users with accounts.
\item Once adversaries obtain the identities of users with accounts and their content sharing histories, they can do anything they wish with that information. 
\end{itemize}
\end{tcolorbox}
\caption{A computer security scenario like Scenario D1 in Figure~\ref{fig:scenario:disclosure:base} except that the company with the vulnerability provides a service that is immoral (though legal) and that harms many individuals. Users who upload content to the company's web service are breaking the law.}
\label{fig:scenario:disclosure:legalthreat:usersbreaklaw}
\end{figure*}

\begin{figure*}
\begin{tcolorbox}
\begin{center}
\textbf{Computer Security Scenario E1: Submission Raises Ethical Concerns (Base Case)}\end{center}
\textbf{Context:}
\begin{itemize}[itemsep=0pt, parsep=2pt]
\item \textbf{Backstory:}
    \begin{itemize}[itemsep=0pt, parsep=2pt]
    \item University researchers discovered a new, serious vulnerability in the software for a webserver that allows remote adversarial (over the Internet) root (administrative-level) access.
    \item The researchers worked with the maintainers of the webserver software to develop and release a patch.
    \item The researchers submitted a paper about their findings to a conference.
    \item Most of the paper focuses on the details of the vulnerability, the discovery method, and the fix; one section (Section 9.3) describes the results of experiments in which the researchers count the number of vulnerable hosts by scanning the IPv4 network address space; the researchers started their Internet-wide scans before they notified the maintainers of the webserver software of the vulnerability; the scans proceeded once a week thereafter; the scans stopped eight weeks after the maintainers of the webserver released their patch.
    \end{itemize}
\item \textbf{On the issue with the paper:}
    \begin{itemize}[itemsep=0pt, parsep=2pt]
    \item The program committee, with expertise in web security and networking, observes that the scanning method in Section 9.3 could have caused some computers --- including insulin pump controllers in hospitals, which have webserver administrative interfaces  --- to crash.
	\item One of the program committee members has contacts at their local hospital and has learned that the hospital's insulin pumps \emph{did} crash every week at exactly the time of the researcher's Internet scans.
    \item After discussion, the program committee reaches consensus: with the knowledge that they have, the scanning experiments described in Section 9.3  should not have been done.
    \item The authors wrote that they contacted their institution’s IRB, which determined that the research was not human subjects research and hence was outside the purview of the IRB.
    \item It is clear that the authors did not know that their scans could cause crashes; however, the authors were negligent or unaware of scanning best practices: they would have known about the potential for crashes (in general, though not necessarily for insulin pump controllers) if they had done more extensive testing of their scanning infrastructure before conducting their full scan.
    \item The program committee believes that it is important to publish the paper because of the scientific quality and contribution of the paper as a whole. The program committee believes that the work described in Section~9.3 is not essential to understand the importance and contributions of the paper as a whole.
    \end{itemize}
\end{itemize}
\textbf{The choice for the program committee (the program chairs have determined that these are the only two options):}
\begin{itemize}[itemsep=0pt, parsep=2pt]
\item \textit{Reject the paper:} The valuable scientific contributions of the paper are not shared with the research community; the research community does not see examples of research that should not have been done (and hence does not internalize that such research is acceptable); authors not rewarded for research that should not have been done.
\item \textit{Accept the paper without any required modifications; authors receive reviews and can choose how to revise their paper:} 
The valuable scientific contributions of the paper are shared with the research community; depending on whether and how the authors choose to revise their paper, the research community can have a public dialog around the ethics of this research; the authors gain another research publication; the authors (and, depending on how the paper is revised, the community) learn that they can publish research that should not have been done. 
\end{itemize}
\end{tcolorbox}
\caption{A computer security scenario in which a program committee determines that a submitted paper includes the results of research that should not have been done.}
\label{fig:scenario:program_committee_concerns:base}
\end{figure*}

\begin{figure*}
\begin{tcolorbox}
\begin{center}
\textbf{Computer Security Scenario E2: Submission Raises Ethical Concerns (Reject, Accept, Remove)}\end{center}
\textbf{Context:} Equivalent to Scenario E1 in Figure~\ref{fig:scenario:program_committee_concerns:base}. \\ 
\textbf{The available choices for the the program committee is now larger:}
\begin{itemize}[itemsep=0pt, parsep=2pt]
\item \textit{Reject the paper:} See Figure~\ref{fig:scenario:program_committee_concerns:base}.
\item \textit{Accept the paper without any required modifications; authors receive reviews and can choose how to revise their paper:}  See Figure~\ref{fig:scenario:program_committee_concerns:base}.
\item \textit{Accept the paper under the condition that the authors agree to remove Section 9.3:} The conference is not seen to have published a paper containing research that should not have been done; the research community does not see examples of research that should not have been done (and hence does not internalize that such research is acceptable); the valuable scientific contributions of the paper (minus Section 9.3) are shared with the research community; the authors receive the benefit of a publication; the authors learn that they can conduct research that should not be done as long as they do not publish the results.
\end{itemize}
\end{tcolorbox}
\caption{A computer security scenario like Scenario E1 in Figure~\ref{fig:scenario:program_committee_concerns:base} except that more options are presented to the program committee.}
\label{fig:scenario:program_committee_concerns:moreoptionsremove}
\end{figure*}

\begin{figure*}
\begin{tcolorbox}
\begin{center}
\textbf{Computer Security Scenario E3: Submission Raises Ethical Concerns (Reject, Accept, Remove, Note)}\end{center}
\textbf{Context:} Equivalent to Scenario E2 in Figure~\ref{fig:scenario:program_committee_concerns:moreoptionsremove}. \\ 
\textbf{The available choices for the the program committee is now larger:}
\begin{itemize}[itemsep=0pt, parsep=2pt]
\item \textit{Reject the paper:} See Figure~\ref{fig:scenario:program_committee_concerns:base}.
\item \textit{Accept the paper without any required modifications; authors receive reviews and can choose how to revise their paper:}  See Figure~\ref{fig:scenario:program_committee_concerns:base}.
\item \textit{Accept the paper under the condition that the authors agree to remove Section 9.3:} See Figure~\ref{fig:scenario:program_committee_concerns:moreoptionsremove}.
\item \textit{Accept the paper under the condition that the authors agree to add a clearly visible note to the first page, written by the program committee; the note will describe the ethical concern and the program committee's belief that the work in Section 9.3 should not have been done:} The conference is seen as having an ethical bar and is known to attach ethics-related notes to papers that contain research that should not have been done; the results of the research (including Section 9.3) are made available to the research community; the research community can have a public dialog around the ethics of this research; the community (and the authors) learn that they can publish research that should not have been done if they are willing to accept an ethics-related note attached by the program committee.
\end{itemize}
\end{tcolorbox}
\caption{A computer security scenario like Scenario E2 in Figure~\ref{fig:scenario:program_committee_concerns:moreoptionsremove} except that more options are presented to the program committee.}
\label{fig:scenario:program_committee_concerns:moreoptions}
\end{figure*}

\begin{figure*}
\begin{tcolorbox}
\begin{center}
\textbf{Computer Security Scenario E4: Submission Raises Ethical Concerns (Extensive Author Testing)}\end{center}
\textbf{Context:} Equivalent to Scenario E3 in Figure~\ref{fig:scenario:program_committee_concerns:moreoptions} except:
\begin{itemize}[itemsep=0pt, parsep=2pt]
\item The program chairs reach out to the authors for additional information, which they share with the program committee, so that the program committee can make a more informed decision.
\item The program chairs learn that the researchers thought that they had done everything possible to prevent crashes and, as a result, believed that no crashes would happen.
\item Among the things that the authors did to assess the potential for crashes:
    \begin{itemize}[itemsep=0pt, parsep=2pt]
    \item Extensive testing within their own networks.
    \item Asking ethics experts in another field (computer networks) to review their scanning infrastructure and methodology before deployment.
	\item The reason that their scanning infrastructure caused crashes on insulin pump computers is due to an unusual property with those insulin pump machines that no one (except for people who work with insulin pumps) would be expected to know about or foresee (the program committee identified the risk because a program committee member has expertise with insulin pump devices).
	\end{itemize}
\item Additionally, the researchers employed mechanisms to detect if crashes happened:
    \begin{itemize}[itemsep=0pt, parsep=2pt]
    \item Setting up a ``complaints'' website hosted at the IP address originating the scans, such that if any network operator had concerns about the scans or a crash did happen, the operators could quickly and easily reach the researchers.
    \item The reason the researchers did not receive feedback via their complaints website was due to procedural errors by the operators of the insulin pumps.
    \end{itemize}
\end{itemize}
\end{tcolorbox}
\caption{A computer security scenario like Scenario E3 in Figure~\ref{fig:scenario:program_committee_concerns:moreoptions} except that the program chairs and program committee learn that the researchers thought that they had done everything possible to prevent crashes from their scans.}
\label{fig:scenario:program_committee_concerns:authorstried}
\end{figure*}

\begin{figure*}
\begin{tcolorbox}
\begin{center}
\textbf{Computer Security Scenario E5: Submission Raises Ethical Concerns (Known Risk and Moral Responsibility)}\end{center}
\textbf{Context:} Equivalent to Scenario E3 in Figure~\ref{fig:scenario:program_committee_concerns:moreoptions} except:
\begin{itemize}[itemsep=0pt, parsep=2pt]
\item The program chairs reach out to the authors for additional information, which they share with the program committee, so that the program committee can make a more informed decision.
\item The program chairs learn that the researchers knew that Internet-wide scans could cause crashes.
\item The researchers proceeded with their Internet-wide scans out of a sense of duty and moral responsibility: given the seriousness of the vulnerability, they believed that it was imperative to conduct their scans, identify still-vulnerable webservers, and contact the operators of those webservers and encourage them to patch.
\item The researchers did not know that the webserver software was installed on insulin pumps; the researchers assumed that all impacts of crashes, if they occurred, would be minimal.
\item The researcher's scans and follow-on efforts to reach operators of vulnerable webservers had a direct and measurable impact on the security of webservers: many webserver operators applied the patch \emph{because} the researchers contacted them.
\item Even though the researchers knew that their scans could cause crashes, they did \textit{not} know that their scans \emph{had} caused crashes. 
\item Due to space limitations, the authors did not discuss their knowledge of the potential for crashes in their submission.
\end{itemize}
\end{tcolorbox}
\caption{A computer security scenario like Scenario E3 in Figure~\ref{fig:scenario:program_committee_concerns:moreoptions} except that the program chairs and program committee learn that the researchers knew that crashes were possible but proceeded anyway due to a sense of duty and moral responsibility.}
\label{fig:scenario:program_committee_concerns:authorsknewandfeltduty}
\end{figure*}

\begin{figure*}
\begin{tcolorbox}
\begin{center}
\textbf{Computer Security Scenario E6: Submission Raises Ethical Concerns (Authors Ignored Risks)}\end{center}
\textbf{Context:} Equivalent to Scenario E3 in Figure~\ref{fig:scenario:program_committee_concerns:moreoptions} except:
\begin{itemize}[itemsep=0pt, parsep=2pt]
\item The program chairs reach out to the authors for additional information, which they share with the program committee, so that the program committee can make a more informed decision.
\item The program chairs learn that the researchers knew that the scans described in Section 9.3 might cause some computers to crash.
\item The program chairs learn that the researchers proceeded with the experiment in Section 9.3 anyway because (1) they believed that the results of their scans in Section 9.3 would increase the likelihood that their paper would be accepted and (2) they reasoned that a few crashes would be fine since computers crash all the time for other reasons, too.
\end{itemize}
\end{tcolorbox}
\caption{A computer security scenario like Scenario E3 in Figure~\ref{fig:scenario:program_committee_concerns:moreoptions} except that the program chairs and program committee learn that the researchers knew that their scans could cause crashes but decided to proceed anyway.}
\label{fig:scenario:program_committee_concerns:authorsknew}
\end{figure*}

\begin{figure*}
\begin{tcolorbox}
\begin{center}
\textbf{Computer Security Scenario E7: Submission Raises Ethical Concerns (Authors Ignored Risks, Luck Prevents Crashes)}\end{center}
\textbf{Context:} Equivalent to Scenario E6 in Figure~\ref{fig:scenario:program_committee_concerns:authorsknew} except:
\begin{itemize}[itemsep=0pt, parsep=2pt]
\item Through sheer luck, the researchers' final crawling infrastructure was poorly implemented, had unnecessary timing delays and, as a result of the timing delays, the researchers' Internet-wide scans did \emph{not} cause any crashes.
\item The results in the paper are otherwise correct --- the bug meant that the crawls were slower than they needed to be, and that the crawls did not cause crashes, but the results in Section 9.3 of the paper are otherwise correct.
\item Based on their preliminary experiments with a crawling tool that did not have the unnecessary delays, the researchers still believed that the scans described in Section 9.3 might cause some computers to crash and chose to proceed with their scans anyway.
\item If the deployed scans did not have the extra (unnecessary) timing delays, the program committee knows that the scans would have caused insulin pump controllers in hospitals to crash.
\end{itemize}
\end{tcolorbox}
\caption{A computer security scenario like Scenario E6 in Figure~\ref{fig:scenario:program_committee_concerns:authorsknew} except that no crashes happened due to sheer luck.}
\label{fig:scenario:program_committee_concerns:authorsknew:nocrashesluck}
\end{figure*}

\begin{figure*}
\begin{tcolorbox}
\begin{center}
\textbf{Computer Security Scenario E8: Submission Raises Ethical Concerns (Authors Ignored Risks, Section 9.3 Results Critical)}\end{center}
\textbf{Context:} Equivalent to Scenario E6 in Figure~\ref{fig:scenario:program_committee_concerns:authorsknew} except:
\begin{itemize}[itemsep=0pt, parsep=2pt]
\item As part of the research underlying Section 9.3, the researchers discovered that many websites remain vulnerable (no configuration change) despite repeated attempts from the researchers to reach the operators of the remaining vulnerable websites. The researchers write about this discovery in Section 9.3.
\item The program committee believes that it is important to publish the results in this paper, including the results in Section 9.3, (1) because of the scientific quality and contribution of the paper as a whole and (2) because the publication of Section 9.3 will provide the final encouragement for website operators to apply the patch; until the patch is applied, the webservers will remain vulnerable to remote root (administrative-level) compromise. Not publishing the paper, including Section 9.3, would thus result in the continued vulnerability of many webservers and hence the continued exposure of many users to harm.
\end{itemize}
\end{tcolorbox}
\caption{A computer security scenario like Scenario E6 in Figure~\ref{fig:scenario:program_committee_concerns:authorsknew} except that there are significant benefits to the world if the paper, including Section 9.3, is published.}
\label{fig:scenario:program_committee_concerns:authorsknew:73critical}
\end{figure*}

\begin{figure*}
\begin{tcolorbox}
\begin{center}
\textbf{Computer Security Scenario E9: Submission Raises Ethical Concerns (Authors Ignored Risks, Moral Implications Not Realized)}\end{center}
\textbf{Context:} 
Equivalent to Scenario E6 in Figure~\ref{fig:scenario:program_committee_concerns:authorsknew} except:
\begin{itemize}[itemsep=0pt, parsep=2pt]
\item After discussing with the researchers, it is clear to the program chairs that the researchers did not realize the moral implications of causing computers to crash; the researchers are now deeply regretful and wish that they had not done the scans discussed in Section 9.3.
\end{itemize}
\end{tcolorbox}
\caption{A computer security scenario like Scenario E6 in Figure~\ref{fig:scenario:program_committee_concerns:authorsknew} except that the program chairs and program committee learn the researchers did not understand the moral implications of their actions and, once they learn of those moral implications, are deeply regretful and wish that they had not done the research described in Section 9.3.}
\label{fig:scenario:program_committee_concerns:authorsknew:phdstudent:implicationsnotrealized}
\end{figure*}

\begin{figure*}
\begin{tcolorbox}
\begin{center}

\textbf{Computer Security Scenario F: Response to Submission Rejection}\end{center}

\textbf{Context:}
\begin{itemize}[itemsep=0pt, parsep=2pt]

\item The authors of the paper in Scenario E1 (Figure~\ref{fig:scenario:program_committee_concerns:base}) receive a rejection; in the rejection email, the program committee explains the ethical concerns with the research underlying Section 9.3 of their submission.
\item The authors did not realize the potential for crashes while doing the research for Section 9.3 and while writing the paper. Now that the concerns have been explained to them, the researchers understand and agree that the network scans reported in Section 9.3 should not have been done.
\item 
The authors believe that the findings in their paper, even without Section 9.3, are important for the field to know.
\item If the authors submit to a different conference  with an entirely different program committee, that new conference and new program committee will not have any communications with the program committee of the conference that already rejected the paper.
\end{itemize}
\textbf{The choice for the researchers:}
\begin{itemize}[itemsep=0pt, parsep=2pt]
\item \textit{Stop working on the paper / project and not submit to any other conference:} The research, in its entirety, is no longer considered by the community.
\item \textit{Submit the paper without modification to a different conference with an entirely different program committee:} A new program committee, composed of different program committee members, will evaluate the paper independently. This new program committee will make an independent decision on whether or not to publish the paper, or whether to ask for modifications to the paper before publication.
\item \textit{Remove Section 9.3 and submit to a different conference with an entirely different program committee:} This new program committee will not see any evidence of the research that should not have been done (the former Section 9.3). Thus, the new program committee will make a decision to publish / not publish the paper only based on the contents of the other sections. 

\item \textit{Add a discussion to Section 9.3 that explains their initial oversights and the fact that their network scans should not have been done and submit to a different conference with an entirely different program committee:} The program committee will know the entirety of the research done, including the research that should not have been done (Section 9.3). It is unknown how the program committee will react to the authors' explanation of the initial oversights and errors with the research that they report in Section 9.3.

\end{itemize}
\end{tcolorbox}
\caption{
A computer security scenario in which researchers receive a rejection  from a conference program committee along with a notice that the program committee believes that part of the research discussed in their submission should not have been done. This scenario follows
Scenario E1 in Figure~\ref{fig:scenario:program_committee_concerns:base}.
}
\label{fig:scenario:resubmit_paper_with_concerns}

\end{figure*}


\begin{thebibliography}{100}

\bibitem{ACM:ethics}
ACM.
\newblock {ACM} code of ethics and professional conduct, 2018.
\newblock {\url{https://www.acm.org/code-of-ethics}}; accessed 06/03/2023.

\bibitem{adar2007user}
Eytan Adar.
\newblock User 4xxxxx9: Anonymizing query logs.
\newblock In {\em Proc of Query Log Analysis Workshop, International Conference
  on World Wide Web}, 2007.

\bibitem{ChineseandWesternersRespondDifferentlytotheTrolleyDilemmas}
Henrik Ahlenius and Torbj{\"o}rn T{\"a}nnsj{\"o}.
\newblock Chinese and {Westerners} respond differently to the trolley dilemmas.
\newblock {\em Journal of Cognition and Culture}, 12(3-4), 2012.

\bibitem{AnscombeJournal}
G.E.M. Anscombe.
\newblock Modern moral philosophy.
\newblock {\em Philosophy}, 33(124), 1958.

\bibitem{NatureQuote}
John~W. Ayers, Theodore~L. Caputi, Camille Nebeker, and Mark Dredze.
\newblock Don't quote me: Reverse identification of research participants in
  social media studies.
\newblock {\em npj Digital Medicine}, 2018.

\bibitem{EthicsToolkit}
Julian Baggini and Peter~S. Fosl.
\newblock {\em The Ethics Toolkit: A Compendium of Ethical Concepts and
  Methods}.
\newblock Blackwell Publishing, 2007.

\bibitem{BeauchampChildress}
Tom~L. Beauchamp and James~F. Childress.
\newblock {\em Principles of Biomedical Ethics}.
\newblock Oxford University Press, 8 edition, 2019.
\newblock First edition published in 1979.

\bibitem{BernsteinESR}
Michael~S. Bernstein, Margaret Levi, David Magnus, Betsy~A. Rajala, Debra Satz,
  and Quinn Waeiss.
\newblock Ethics and society review: Ethics reflection as a precondition to
  research funding.
\newblock {\em PNAS}, 118(52), 2021.

\bibitem{blaze1994protocol}
Matt Blaze.
\newblock Protocol failure in the escrowed encryption standard.
\newblock In {\em ACM CCS}, 1994.

\bibitem{BlazeLocks}
Matt Blaze.
\newblock Rights amplification in master-keyed mechanical locks.
\newblock {\em IEEE Security {\&} Privacy}, 1(2), 2003.

\bibitem{BlazeTweet}
Matt Blaze, July 2022.
\newblock {\url{https://twitter.com/mattblaze/status/1553244850304290817}};
  accessed 01/28/2023.

\bibitem{bruckman2002studying}
Amy Bruckman.
\newblock Studying the amateur artist: A perspective on disguising data
  collected in human subjects research on the {I}nternet.
\newblock {\em Ethics and Information Technology}, 4(3), 2002.

\bibitem{buolamwini2018gender}
Joy Buolamwini and Timnit Gebru.
\newblock Gender shades: Intersectional accuracy disparities in commercial
  gender classification.
\newblock In {\em Conference on Fairness, Accountability and Transparency},
  2018.

\bibitem{burgess2022watching}
Ben Burgess, Avi Ginsberg, Edward~W Felten, and Shaanan Cohney.
\newblock Watching the watchers: Bias and vulnerability in remote proctoring
  software.
\newblock In {\em USENIX Security}, 2022.

\bibitem{caliskan2017semantics}
Aylin Caliskan, Joanna~J. Bryson, and Arvind Narayanan.
\newblock Semantics derived automatically from language corpora contain
  human-like biases.
\newblock {\em Science}, 356(6334), 2017.

\bibitem{COSregisteredreports}
{Center for Open Science}, 2023.
\newblock {\url{https://www.cos.io/initiatives/registered-reports}}; accessed
  06/04/2023.

\bibitem{chatterjee2018spyware}
Rahul Chatterjee, Periwinkle Doerfler, Hadas Orgad, Sam Havron, Jackeline
  Palmer, Diana Freed, Karen Levy, Nicola Dell, Damon McCoy, and Thomas
  Ristenpart.
\newblock The spyware used in intimate partner violence.
\newblock In {\em IEEE Symposium on Security and Privacy}, 2018.

\bibitem{chivukula2021surveying}
Shruthi~Sai Chivukula, Ziqing Li, Anne~C Pivonka, Jingning Chen, and Colin~M
  Gray.
\newblock Surveying the landscape of ethics-focused design methods.
\newblock {\em arXiv preprint arXiv:2102.08909}, 2021.

\bibitem{AcademiesCrypto}
{Committee on Law Enforcement and Intelligence Access to Plaintext
  Information}.
\newblock {\em Decrypting the Encryption Debate: A Framework for Decision
  Makers}.
\newblock National Academies of Sciences, Engineering, and Medicine, 2018.

\bibitem{conway2013deontological}
Paul Conway and Bertram Gawronski.
\newblock Deontological and utilitarian inclinations in moral decision making:
  A process dissociation approach.
\newblock {\em Journal of Personality and Social Psychology}, 104(2), 2013.

\bibitem{conway2018sacrificial}
Paul Conway, Jacob Goldstein-Greenwood, David Polacek, and Joshua~D. Greene.
\newblock Sacrificial utilitarian judgments do reflect concern for the greater
  good: Clarification via process dissociation and the judgments of
  philosophers.
\newblock {\em Cognition}, 179, 2018.

\bibitem{daffalla2021defensive}
Alaa Daffalla, Lucy Simko, Tadayoshi Kohno, and Alexandru~G Bardas.
\newblock Defensive technology use by political activists during the {Sudanese}
  revolution.
\newblock In {\em IEEE Symposium on Security and Privacy}, 2021.

\bibitem{d2022ethics}
Alexandra D'Arcy and Emily~M Bender.
\newblock Ethics in linguistics.
\newblock {\em Annual Review of Linguistics}, 9, 2022.

\bibitem{DeighIntroEthics}
John Deigh.
\newblock {\em An Introduction to Ethics}.
\newblock Cambridge University Press, 2010.

\bibitem{FreeHaven}
Roger Dingledine.
\newblock The free haven project: Design and deployment of an anonymous secure
  data haven.
\newblock Masters thesis, Massachusetts Institute of Technology, 2000.

\bibitem{Tor}
Roger Dingledine, Nick Mathewson, and Paul Syverson.
\newblock Tor: The second-generation onion router.
\newblock In {\em USENIX Security}, 2004.

\bibitem{BotnetEthicsProblems}
David Dittrich, Felix Leder, and Tillmann Werner.
\newblock A case study in ethical decision making regarding remote mitigation
  of botnets.
\newblock In {\em Financial Cryptography and Data Security}, 2010.

\bibitem{FDAMITREPlaybook2018}
Fred Donovan.
\newblock {FDA} unveils {MITRE’s} medical device security playbook, 2018.
\newblock
  {\url{https://healthitsecurity.com/news/fda-unveils-mitres-medical-device-security-playbook}};
  accessed 12/26/2022.

\bibitem{DriverEthics}
Julia Driver.
\newblock {\em Ethics: The Fundamentals}.
\newblock Blackwell Publishing, 2006.

\bibitem{durumeric2013zmap}
Zakir Durumeric, Eric Wustrow, and J~Alex Halderman.
\newblock {ZMap}: Fast internet-wide scanning and its security applications.
\newblock In {\em USENIX Security}, 2013.

\bibitem{FieslerFan}
Brianna Dym and Casey Fiesler.
\newblock Ethical and privacy considerations for research using online fandom
  data.
\newblock {\em Fan Studies Methodologies}, 33, 2020.

\bibitem{evtimov2021foggysight}
Ivan Evtimov, Pascal Sturmfels, and Tadayoshi Kohno.
\newblock Foggysight: A scheme for facial lookup privacy.
\newblock {\em Proceedings on Privacy Enhancing Technologies}, 2021.

\bibitem{field2021survey}
Anjalie Field, Su~Lin Blodgett, Zeerak Waseem, and Yulia Tsvetkov.
\newblock A survey of race, racism, and anti-racism in {NLP}.
\newblock {\em arXiv preprint arXiv:2106.11410}, 2021.

\bibitem{FieslerTwitter}
Casey Fiesler and Nicholas Proferes.
\newblock ``{P}articipant'' perceptions of {T}witter research ethics.
\newblock {\em Social Media + Society}, 4(1), 2018.

\bibitem{FloridiHandbook}
Luciano Floridi.
\newblock {\em The Cambridge Handbook of Information and Computer Ethics}.
\newblock Cambridge University Press, 2010.

\bibitem{Foot2002MoralDilemmas}
Philippa Foot.
\newblock {\em Moral Dilemmas: And Other Topics in Moral Philosophy}.
\newblock Oxford University Press UK, 2002.

\bibitem{freed2017digital}
Diana Freed, Jackeline Palmer, Diana Minchala, Karen Levy, Thomas Ristenpart,
  and Nicola Dell.
\newblock Digital technologies and intimate partner violence: {A} qualitative
  analysis with multiple stakeholders.
\newblock {\em Proceedings of the ACM on Human-Computer Interaction}, 1(CSCW),
  2017.

\bibitem{freed2018stalker}
Diana Freed, Jackeline Palmer, Diana Minchala, Karen Levy, Thomas Ristenpart,
  and Nicola Dell.
\newblock ``{A} {S}talker's {P}aradise'': {H}ow intimate partner abusers
  exploit technology.
\newblock In {\em ACM CHI}, 2018.

\bibitem{friedman2019value}
Batya Friedman and David~G Hendry.
\newblock {\em Value Sensitive Design: Shaping Technology with Moral
  Imagination}.
\newblock The MIT Press, 2019.

\bibitem{friedman2007human}
Batya Friedman and Peter~H Kahn~Jr.
\newblock Human values, ethics, and design.
\newblock In {\em The Human-Computer Interaction Handbook}. CRC press, 2007.

\bibitem{friedman2006development}
Batya Friedman, Ian Smith, Peter H.~Kahn, Sunny Consolvo, and Jaina Selawski.
\newblock Development of a privacy addendum for open source licenses: {Value
  Sensitive Design} in industry.
\newblock In {\em International Conference on Ubiquitous Computing}, 2006.

\bibitem{SimsonWebSecurity}
Simson Garfinkel.
\newblock {\em Web Security, Privacy {\&} Commerce}.
\newblock O'Reilly Media, Inc., 2 edition, 2001.

\bibitem{sep-morality-definition}
Bernard Gert and Joshua Gert.
\newblock The definition of morality.
\newblock In Edward~N. Zalta, editor, {\em The {Stanford} Encyclopedia of
  Philosophy}. Metaphysics Research Lab, Stanford University, {F}all 2020
  edition, 2020.

\bibitem{geuss1981idea}
Raymond Geuss et~al.
\newblock {\em The Idea of a Critical Theory: {Habermas} and the {Frankfurt}
  School}.
\newblock Cambridge University Press, 1981.

\bibitem{gold_colman_pulford_2014}
Natalie Gold, Andrew~M. Colman, and Briony~D. Pulford.
\newblock Cultural differences in responses to real-life and hypothetical
  trolley problems.
\newblock {\em Judgment and Decision Making}, 9(1), 2014.

\bibitem{habermas1994justification}
J.~Habermas and C.~Cronin.
\newblock {\em Justification and Application: Remarks on Discourse Ethics}.
\newblock Studies in Contemporary German. MIT Press, 1994.

\bibitem{HalHeyRan2008ICD}
Daniel Halperin, Thomas~S. Heydt-Benjamin, Benjamin Ransford, Shane~S. Clark,
  Benessa Defend, Will Morgan, Kevin Fu, Tadayoshi Kohno, and William~H.
  Maisel.
\newblock Pacemakers and implantable cardiac defibrillators: Software radio
  attacks and zero-power defenses.
\newblock In {\em IEEE Symposium on Security and Privacy}, May 2008.

\bibitem{hare1981moral}
Richard~Mervyn Hare.
\newblock {\em Moral thinking: Its levels, Method, and Point}.
\newblock Oxford: Clarendon Press; New York: Oxford University Press, 1981.

\bibitem{havron2019clinical}
Sam Havron, Diana Freed, Rahul Chatterjee, Damon McCoy, Nicola Dell, and Thomas
  Ristenpart.
\newblock Clinical computer security for victims of intimate partner violence.
\newblock In {\em {USENIX} Security}, 2019.

\bibitem{holz2021ieee}
Thorsten Holz and Alina Oprea.
\newblock {IEEE S\&P'21} program committee statement regarding the ``{Hypocrite
  Commits}'' paper, 2021.
\newblock
  {\url{https://www.ieee-security.org/TC/SP2021/downloads/2021_PC_Statement.pdf}};
  accessed 02/04/2023.

\bibitem{holz2008measurements}
Thorsten Holz, Moritz Steiner, Frederic Dahl, Ernst~W Biersack, Felix~C
  Freiling, et~al.
\newblock Measurements and mitigation of peer-to-peer-based botnets: A case
  study on storm worm.
\newblock In {\em Leet}, 2008.

\bibitem{honneth2014freedom}
Axel Honneth.
\newblock {\em Freedom's Right: The Social Foundations of Democratic Life}.
\newblock Columbia University Press, 2014.

\bibitem{horkheimer1937traditional}
Max Horkheimer.
\newblock {\em Critical Theory: Selected Essays}.
\newblock Continuum Publishing Corporation, 1975.
\newblock Originally published in 1937.

\bibitem{horkheimer1947theodor}
Max Horkheimer and Theodor~W. Adorno.
\newblock Dialektik der aufkl{\"a}rung.
\newblock {\em Philosophische Fragmente}, 14, 1947.

\bibitem{IEEE:ethics}
IEEE.
\newblock {IEEE} code of ethics, 2020.
\newblock {\url{https://www.ieee.org/about/corporate/governance/p7-8.html}};
  accessed 06/03/2023.

\bibitem{Oakland2022CFP}
{IEEE Computer Society Technical Committee on Security and Privacy}, 2021.
\newblock {\url{https://www.ieee-security.org/TC/SP2022/cfpapers.html}};
  accessed 01/29/2023.

\bibitem{Oakland2023CFP}
{IEEE Computer Society Technical Committee on Security and Privacy}, 2022.
\newblock {\url{https://www.ieee-security.org/TC/SP2023/cfpapers.html}};
  accessed 01/29/2023.

\bibitem{IphofenHandbook}
Ron Iphofen.
\newblock {\em Handbook of Research Ethics and Scientific Integrity}.
\newblock Springer, 2020.

\bibitem{AcademiesSpectre}
Anne Johnson and Lynette~I. Millett, editors.
\newblock {\em Beyond Spectre: Confronting New Technical and Policy Challenges:
  Proceedings of a Workshop}.
\newblock National Academies of Sciences, Engineering, and Medicine, 2019.

\bibitem{ethicscensorship}
Ben Jones, Roya Ensafi, Nick Feamster, Vern Paxson, and Nick Weaver.
\newblock Ethical concerns for censorship measurement.
\newblock In {\em Proceedings of the 2015 ACM SIGCOMM Workshop on Ethics in
  Networked Systems Research}, 2015.

\bibitem{VWarticle}
Tushar Kamath.
\newblock {VW} spends years trying to conceal security flaw, 2015.
\newblock
  {\url{https://www.team-bhp.com/news/vw-spends-years-trying-conceal-security-flaw}};
  accessed 1/14/2023.

\bibitem{Spamalytics}
Chris Kanich, Christian Kreibich, Kirill Levchenko, Brandon Enright,
  Geoffrey~M. Voelker, Vern Paxson, and Stefan Savage.
\newblock Spamalytics: An empirical analysis of spam marketing conversion.
\newblock In {\em ACM CCS}, 2008.

\bibitem{kant1785grundlegung}
Immanuel Kant.
\newblock Grundlegung zur metaphysik der sitten [groundwork of the metaphysics
  of morals].
\newblock {\em Riga, Latvia: JF Hartknoch}, 1785.

\bibitem{SecurityEthicsConference}
Tadayoshi Kohno, Yasemin Acar, and Wulf Loh.
\newblock Ethical frameworks and computer security trolley problems:
  Foundations for conversations.
\newblock In {\em USENIX Security}, 2023.

\bibitem{FDAadvisoryIMD}
Daniel~B. Kramer and Kevin Fu.
\newblock Cybersecurity concerns and medical devices: Lessons from a pacemaker
  advisory.
\newblock {\em JAMA}, 318(21), 2017.

\bibitem{lamont2017bridging}
Mich{\`e}le Lamont, Laura Adler, Bo~Yun Park, and Xin Xiang.
\newblock Bridging cultural sociology and cognitive psychology in three
  contemporary research programmes.
\newblock {\em Nature Human Behaviour}, 1(12), 2017.

\bibitem{CryptoWars}
Steven Levy.
\newblock {\em Crypto: How the Code Rebels Beat the Government Saving Privacy
  in the Digital Age}.
\newblock Penguin Books, 2002.

\bibitem{locke1988}
John Locke.
\newblock {\em Locke: Two Treatises of Government}.
\newblock Cambridge University Press, 1988.
\newblock Originally published in 1689.

\bibitem{manzeschke2015meestar}
Arne Manzeschke, Karsten Weber, Elisabeth Rother, and Heiner Fangerau.
\newblock {\em Ethical Questions in the Area of Age Appropriate Assisting
  Systems}.
\newblock 2015.

\bibitem{MarkHam2012}
Annette Markham.
\newblock Fabrication as ethical practice.
\newblock {\em Information, Communication \& Society}, 15(3), 2012.

\bibitem{matthews2017stories}
Tara Matthews, Kathleen O'Leary, Anna Turner, Manya Sleeper, Jill~Palzkill
  Woelfer, Martin Shelton, Cori Manthorne, Elizabeth~F. Churchill, and Sunny
  Consolvo.
\newblock Stories from survivors: {P}rivacy \& security practices when coping
  with intimate partner abuse.
\newblock In {\em ACM CHI}, 2017.

\bibitem{KatieBlog}
Katie Moussouris.
\newblock Coordinated vulnerability disclosure: Bringing balance to the force,
  July 2010.
\newblock
  {\url{https://learn.microsoft.com/en-us/archive/blogs/ecostrat/coordinated-vulnerability-disclosure-bringing-balance-to-the-force}};
  accessed 01/28/2023.

\bibitem{KatieTweet}
Katie Moussouris, February 2020.
\newblock {\url{https://twitter.com/k8em0/status/1233075185734959104}};
  accessed 01/28/2023.

\bibitem{netflixprize}
Arvind Narayanan and Vitaly Shmatikov.
\newblock Robust de-anonymization of large sparse datasets.
\newblock In {\em IEEE Symposium on Security and Privacy}, 2008.

\bibitem{NSPE:ethics}
NSPE.
\newblock {NSPE} code of ethics for engineers, 2019.
\newblock {\url{https://www.nspe.org/resources/ethics/code-ethics/}}; accessed
  06/03/2023.

\bibitem{parfit1984reasons}
Derek Parfit.
\newblock {\em Reasons and Persons}.
\newblock OUP Oxford, 1984.

\bibitem{FieslerReddit}
Nicholas Proferes, Naiyan Jones, Sarah Gilbert, Casey Fiesler, and Michael
  Zimmer.
\newblock Studying {R}eddit: A systematic overview of disciplines, approaches,
  methods, and ethics.
\newblock {\em Social Media + Society}, 7(2), 2021.

\bibitem{puett2016path}
Michael Puett and Christine Gross-Loh.
\newblock {\em The Path: What Chinese Philosophers Can Teach Us About the Good
  Life}.
\newblock Simon and Schuster, 2016.

\bibitem{QuinnEthics}
Michael~J. Quinn.
\newblock {\em Ethics for the Information Age}.
\newblock Pearson, 7 edition, 2017.

\bibitem{raji2020saving}
Inioluwa~Deborah Raji, Timnit Gebru, Margaret Mitchell, Joy Buolamwini,
  Joonseok Lee, and Emily Denton.
\newblock Saving face: Investigating the ethical concerns of facial recognition
  auditing.
\newblock In {\em AAAI/ACM AIES}, 2020.

\bibitem{IMDguide}
Benjamin Ransford, Daniel~B. Kramer, Denis~Foo Kune, Julio~Auto de~Medeiros,
  Chen Yan, Wenyuan Xu, Thomas Crawford, and Kevin Fu.
\newblock Cybersecurity and medical devices: A practical guide for cardiac
  electrophysiologists.
\newblock {\em Pacing and Clinical Electrophysiology}, 40(8), 2017.

\bibitem{rogaway2015moral}
Phillip Rogaway.
\newblock The moral character of cryptographic work.
\newblock {\em Cryptology ePrint Archive}, 2015.

\bibitem{Justice}
Michael~J. Sandel.
\newblock {\em Justice: What's the Right Thing to Do?}
\newblock Farrar, Straus and Giroux, 2010.

\bibitem{scanlon2000we}
Thomas~M Scanlon.
\newblock {\em What We Owe to Each Other}.
\newblock Harvard University Press, 2000.

\bibitem{StuartSurvey2}
Stuart Schechter and Cristian Bravo-Lillo.
\newblock Ethical-response survey report: Fall 2014, October 2014.
\newblock
  {\url{https://www.microsoft.com/en-us/research/wp-content/uploads/2016/02/EthicalResponse.pdf}};
  accessed 01/29/2023.

\bibitem{StuartSurvey1}
Stuart Schechter and Cristian Bravo-Lillo.
\newblock Using ethical-response surveys to identify sources of disapproval and
  concern with {Facebook’s} emotional contagion experiment and other
  controversial studies, October 2014.
\newblock
  {\url{https://www.microsoft.com/en-us/research/wp-content/uploads/2016/02/Ethical-Response20Survey202014-10-30.pdf}};
  accessed 01/29/2023.

\bibitem{shan2020fawkes}
Shawn Shan, Emily Wenger, Jiayun Zhang, Huiying Li, Haitao Zheng, and Ben~Y
  Zhao.
\newblock Fawkes: Protecting privacy against unauthorized deep learning models.
\newblock In {\em USENIX Security}, 2020.

\bibitem{NetworkLegal}
Douglas~C. Sicker, Paul Ohm, and Dirk Grunwald.
\newblock Legal issues surrounding monitoring during network research.
\newblock In {\em IMC}, 2007.

\bibitem{NumberIMDs}
Barbara~G. Silverman, Thomas~P. Gross, Ronald~G. Kaczmarek, Peggy Hamilton, and
  Stanford Hamburger.
\newblock The epidemiology of pacemaker implantation in the {United States}.
\newblock {\em Public Health Reports}, 110(1), 1995.

\bibitem{StanfordEthics}
{Stanford University, Philosophy Department, Metaphysics Research Lab}.
\newblock Stanford encyclopedia of philosophy, 2022.
\newblock {\url{https://plato.stanford.edu/index.html}}; accessed 12/29/2022.
  Principal editors: Edward N. Zalta, Uri Nodelman; associate editors: Colin
  Allen, Hannah Kim, Paul Oppenheimer; assistant editors: Emma Pease, Lauren
  Thomas, Jesse Alama.

\bibitem{UCSB:Torpig}
Brett Stone-Gross, Marco Cova, Lorenzo Cavallaro, Bob Gilbert, Martin
  Szydlowski, Richard Kemmerer, Christopher Kruegel, and Giovanni Vigna.
\newblock Your botnet is my botnet: Analysis of a botnet takeover.
\newblock In {\em ACM CCS}, 2009.

\bibitem{IEEESpectrumEye}
Eliza Strickland and Mark Harris.
\newblock Their bionic eyes are now obsolete and unsupported.
\newblock {\em IEEE Spectrum}, February 2022.

\bibitem{stuart1863utilitarianism}
John Stuart~Mill.
\newblock Utilitarianism.
\newblock {\em Parker, Son, and Bourn, London}, 1863.

\bibitem{sweenymerge}
Latanya Sweeney.
\newblock Weaving technology and policy together to maintain confidentiality.
\newblock {\em Journal of Law, Medicine and Ethics}, 25(2--3), 1977.

\bibitem{thomas2017ethical}
Daniel~R Thomas, Sergio Pastrana, Alice Hutchings, Richard Clayton, and
  Alastair~R Beresford.
\newblock Ethical issues in research using datasets of illicit origin.
\newblock In {\em IMC}, 2017.

\bibitem{EmilyCare}
Emily Tseng, Mehrnaz Sabet, Rosanna Bellini, Harkiran~Kaur Sodhi, Thomas
  Ristenpart, and Nicola Dell.
\newblock Care infrastructures for digital security in intimate partner
  violence.
\newblock In {\em ACM CHI}, 2022.

\bibitem{Census2022}
{U.S. Department of Commerce}.
\newblock {U.S.} population estimated at 332,403,650 on {Jan.\ 1}, 2022,
  January 2022.
\newblock
  {\url{https://www.commerce.gov/news/blog/2022/01/us-population-estimated-332403650-jan-1-2022};
  accessed 12/26/2022}.

\bibitem{Belmont}
{U.S.\ Department of Health, Education, and Welfare}.
\newblock The {Belmont} report: Ethical principles and guidelines for the
  protection of human subjects of research, April 1979.
\newblock
  \url{https://www.hhs.gov/ohrp/regulations-and-policy/belmont-report/read-the-belmont-report/index.html};
  accessed 12/29/2022. Members of commission: Kenneth John Ryan, Joseph V.
  Brady, Robert E. Cooke, Dorothy I. Height, Albert R. Jonsen, Patricia King,
  Karen Lebacqz, David W. Louisell, Donald W. Seldin, Eliot Stellar, Robert H.
  Turtle.

\bibitem{FDAsecurityWeb}
{U.S.\ Food and Drug Administration}.
\newblock Cybersecurity, 2022.
\newblock
  {\url{https://www.fda.gov/medical-devices/digital-health-center-excellence/cybersecurity}};
  accessed 12/26/2022.

\bibitem{Menlo}
{U.S.\ Homeland Security}.
\newblock The {Menlo} report: Ethical principles guiding information and
  communication technology research, August 2012.
\newblock
  \url{https://www.dhs.gov/sites/default/files/publications/CSD-MenloPrinciplesCORE-20120803_1.pdf};
  accessed 12/29/2022. Working group participants: Michael Bailey, Aaron
  Burstein, KC Claffy, Shari Clayman, David Dittrich (Co-Lead), John Heidemann,
  Erin Kenneally (Co-Lead), Douglas Maughan, Jenny McNeill, Peter Neumann,
  Charlotte Scheper, Lee Tien, Christos Papadopoulos, Wendy Visscher, Jody
  Westby.

\bibitem{USENIXSecurity2023CFP}
{USENIX Association}, 2022.
\newblock
  {\url{https://www.usenix.org/sites/default/files/sec23_cfp_010423.pdf}};
  accessed 01/29/2023.

\bibitem{SCU:Ethics}
Shannon Vallor, Irina Raicu, and Brian Green.
\newblock Technology and engineering practice: Ethical lenses to look through,
  2020.
\newblock
  {\url{https://www.scu.edu/ethics-in-technology-practice/ethical-lenses/}};
  accessed 01/16/2023; from Ethics in Technology Practice, the Markkula Center
  for Applied Ethics at Santa Clara University,
  \url{https://www.scu.edu/ethics/}.

\bibitem{Megamos}
Roel Verdult, Flavio~D. Garcia, and Barıs¸ Ege.
\newblock Dismantling {Megamos} crypto: Wirelessly lockpicking a vehicle
  immobilizer.
\newblock In {\em Supplement to the Proceedings of the 22nd USENIX Security
  Symposium}, 2013.
\newblock
  \url{https://www.usenix.org/sites/default/files/sec15_supplement.pdf}.

\bibitem{techreviewAIbias}
Sheridan Wall and Hilke Schellmann.
\newblock {LinkedIn's} job-matching {AI} was biased. {T}he company’s
  solution? {M}ore {AI}, June 2021.
\newblock
  {\url{https://www.technologyreview.com/2021/06/23/1026825/linkedin-ai-bias-ziprecruiter-monster-artificial-intelligence/}};
  accessed 01/22/2023.

\bibitem{yamamoto2020causes}
Shoko Yamamoto and Masaki Yuki.
\newblock What causes cross-cultural differences in reactions to the trolley
  problem? a cross-cultural study on the roles of relational mobility and
  reputation expectation.
\newblock {\em The Japanese Journal of Social Psychology}, 2020.

\bibitem{ethicsinsecurityresearch}
Yiming Zhang, Mingxuan Liu, Mingming Zhang, Chaoyi Lu, and Haixin Duan.
\newblock Ethics in security research: Visions, reality, and paths forward.
\newblock In {\em 2022 IEEE European Symposium on Security and Privacy
  Workshops (EuroS{\&}PW)}, 2022.

\bibitem{ZimmerPublic}
Michael Zimmer.
\newblock Addressing conceptual gaps in big data research ethics: An
  application of contextual integrity.
\newblock {\em Social Media + Society}, 4(2), 2018.

\bibitem{PhilZimmermannLetter}
Phil Zimmermann.
\newblock Why do you need {PGP}?
\newblock {\em The Ethical Spectacle}, 7 1995.

\end{thebibliography}
\end{document}